\documentclass[journal, draftcls, onecolumn]{IEEEtran}
\usepackage[utf8]{inputenc} 
\usepackage[T1]{fontenc}
\usepackage{url}
\usepackage{ifthen}
\usepackage{cite}
\usepackage[cmex10]{amsmath} 
\usepackage{amssymb,mathrsfs}
\usepackage{bm}
\usepackage{bbm}
\usepackage{graphicx}
\usepackage{color}
\usepackage{xcolor}
\usepackage{subfig}
\usepackage{hyperref}
\hypersetup{colorlinks,breaklinks,
            linkcolor=[RGB]{0,0,205},
            citecolor=[RGB]{34,139,34},
            urlcolor=[RGB]{0,0,128}
            }

\usepackage[normalem]{ulem}

\newtheorem{lemma}{Lemma}[section]
\newtheorem{theorem}{Theorem}[section]

\newtheorem{definition}{Definition}[section]

\newcommand{\CQGT}[1]{\textsf{$(#1)$-CQGT}}
\newcommand{\SCQGT}[1]{\textsf{$(#1)$-SCQGT}}

\newcommand{\aaaa}{\mathrm{(a)}}
\newcommand{\bbbb}{\mathrm{(b)}}
\newcommand{\cccc}{\mathrm{(c)}}
\newcommand{\dddd}{\mathrm{(d)}}
\newcommand{\eeee}{\mathrm{(e)}}

\newcommand{\lp}{\left(}
\newcommand{\rp}{\right)}
\newcommand{\lb}{\left[}
\newcommand{\rb}{\right]}
\newcommand{\lbp}{\left\{}
\newcommand{\rbp}{\right\}}
\newcommand{\lba}{\left\lvert}
\newcommand{\rba}{\right\rvert}
\newcommand{\lnorm}{\left\lVert}
\newcommand{\rnorm}{\right\rVert}
\newcommand{\mv}{\middle\vert}

\newcommand{\mcal}{\mathcal}

\newcommand{\mb}{\mathbf}
\newcommand{\mbb}{\mathbb}
\newcommand{\msf}{\mathsf}

\newcommand{\ra}{\rightarrow}

\newcommand{\slceil}{\lvert\!\lceil}
\newcommand{\srceil}{\rceil\!\rvert}
\newcommand{\hlceil}{\lVert\!\lceil}
\newcommand{\hrceil}{\rceil\!\rVert}

\newcommand{\lfl}{\left\lfloor}
\newcommand{\rfl}{\right\rfloor}

\newcommand{\diid}{\overset{\text{i.i.d.}}{\sim}}

\newcommand{\e}{\mathrm{e}}

\newcommand{\Ber}{\mathrm{Ber}}
\newcommand{\Unif}{\mathrm{Unif}}

\allowdisplaybreaks

\begin{document}
%\title{On the Noisy Coin Weighing Problem} 
%\title{On the Noisy Combinatorial Quantitative Group Testing Problem}
%\title{On the Combinatorial Quantitative Group Testing Problem with Noisy Query and Partial Recovery}
\title{Non-adaptive Combinatorial Quantitative Group Testing with Adversarially Perturbed Measurements}

\author{Yun-Han Li and I-Hsiang Wang
\thanks{The material in this paper was presented in part at the 2020 IEEE Information Theory Workshop (ITW), April 2021 \cite{LiWang_21}.}
\thanks{Y.-H. Li is with the Graduate Institute of Communication Engineering, 
National Taiwan University, Taipei 10617, Taiwan (email: r07942058@ntu.edu.tw).}% <-this % stops a space
\thanks{I.-H. Wang is with the Department of Electrical Engineering and the Graduate Institute of Communication Engineering, National Taiwan University, Taipei 10617, Taiwan (email: ihwang@ntu.edu.tw).}
}

\maketitle

%%%%%%
%% Abstract: 
%% If your paper is eligible for the student paper award, please add
%% the comment "THIS PAPER IS ELIGIBLE FOR THE STUDENT PAPER
%% AWARD." as a first line in the abstract. 
%% For the final version of the accepted paper, please do not forget
%% to remove this comment!
%%
\begin{abstract}
In this paper, combinatorial quantitative group testing (QGT) with noisy measurements is studied. The goal of QGT is to detect defective items from a data set of size $n$ with counting measurements, each of which counts the number of defects in a selected pool of items. 
While most literatures consider either probabilistic QGT with random noise or combinatorial QGT with noiseless measurements, our focus is on the 
combinatorial QGT with noisy measurements that might be adversarially perturbed by additive bounded noises. Since perfect detection is impossible, a partial detection criterion is adopted. With the adversarial noise being bounded by $d_n = \Theta(n^\delta)$ and the detection criterion being to ensure no more than $k_n = \Theta(n^\kappa)$ errors can be made, our goal is to characterize the fundamental limit on the number of measurement, termed \emph{pooling complexity}, as well as provide explicit construction of measurement plans with optimal pooling complexity and efficient decoding algorithms.  
We first show that the fundamental limit is $\frac{1}{1-2\delta}\frac{n}{\log n}$ to within a constant factor not depending on $(n,\kappa,\delta)$ for the non-adaptive setting when $0<2\delta\leq \kappa <1$, sharpening the previous result by Chen and Wang \cite{8006977}. 
We also provide an explicit construction of a non-adaptive deterministic measurement plan with $\frac{1}{1-2\delta}\frac{n}{\log_{2} n}$ pooling complexity up to a constant factor, matching the fundamental limit, with decoding complexity being $o(n^{1+\rho})$ for all $\rho > 0$, nearly linear in $n$, the size of the data set. Our construction combines Hadamard matrices and lossless unbalanced bipartite expanders based on a novel embedding approach.
%%%%%%%%%%%%%%%
%%%%%%%%%%%%%%%
%%%%%%%%%%%%%%%
%$O\lp n\frac{\lambda\lp\lambda-2\delta\rp \log(n)^{3}}{\epsilon}2^{\sqrt{\frac{\lambda \log(n) \log\lp \frac{\lambda \log(n)^{2}}{\epsilon}\rp }{3}}}\rp$.
%is shown to be $\frac{1}{1-2\delta}\frac{n}{\log(n)}$ to within a constant factor not depending on $(n,\kappa,\delta)$ for the non-adaptive setting.

\end{abstract}
%\noindent\textit{An extended version of this paper is accessible at:}\\
%\centerline{\url{http://homepage.ntu.edu.tw/~ihwang/Eprint/itw20cqgt.pdf}}

%% The paper must be self-contained. However, if you are referring to
%% a full version for checking certain proofs, please provide the
%% publicly accessible location below.  If the paper is completely
%% self-contained, you can remove the following line from your
%% submission.
%\textit{A full version of this paper is accessible at:}
%\url{http://isit2019.fr/} 

\section{Introduction}\label{sec:intro}

Group testing is the problem of identifying defective items in a large set with cardinality $n$ by taking measurements on pools (subsets) of items. The type of measurement plays a central role in the fundamental limits of detection efficiency. In a classical model by Dorfman \cite{dorfman1943detection}, binary-valued measurements are considered, where the output is a bit indicating the existence of defected items in the measured pool. Extensive results for this model (termed traditional group testing hereafter), including algorithms and information theoretic limits, can be found in surveys \cite{du1993combinatorial,CIT099} and the references therein. 
Meanwhile, in many modern applications such as bioinformatics \cite{CaoLi_14}, network traffic monitoring \cite{7308980}, resource allocation in multi-user communication systems \cite{10.1007/978-3-030-25027-0_10}, etc., 
more informative measurement on the pool of items can be carried out. 
A natural one is the counting measurement that outputs the number of defective items in the pool. 
% and hence the information revealed in each measurement scales with size of the data set logarithmically. 
This is called the \emph{quantitative group testing} (QGT) problem or the \emph{coin weighing problem} with its root in combinatorics dating back to Shapiro \cite{ShapiroFine_60}. QGT with noiseless measurements has been extensively studied. In particular, it has been shown that the minimum number of measurements is asymptotically $\frac{2n}{\log_2 n}$ \cite{Erdgs2001ONTP,lindstrom_1965,cantor_mills_1966}. 
%in \cite{Erdgs2001ONTP}, and explicit constructions of the optimal non-adaptive measurement plans are given in \cite{lindstrom_1965,cantor_mills_1966}. 
These results are combinatorial in nature as the goal is to detect the defects no matter where they are located. Hence, it is also called the \emph{combinatorial QGT} (CQGT) with noiseless measurements, to contrast another more recent line of works taking a probabilistic approach \cite{AlaouiRamdas_19,sparse_graph_code}, termed \emph{probabilistic QGT} hereafter.

In practice, however, measurement might be noisy, as counting the number of defectives might be too costly to be accurate. In database applications, in order to preserve privacy, the measurement might also be perturbed intentionally \cite{Dwork_06}. While the traditional group testing with noisy measurements has been extensively studied (see \cite{CIT099} for a survey), QGT with noisy measurements is far less understood. 
One line of works pertains to probabilistic QGT with \emph{random} perturbation in the measurement \cite{scarlett2017phase}. 
% the probabilistic setting where there is a prior distribution imposed on the location of the defectives and the perturbation is \emph{random} \cite{scarlett2017phase}. 
% and aims to detect the defective items to a certain degree with a target probability \cite{scarlett2017phase}. 
Another line of works \cite{Dinur2003RevealingIW,dwork2008new,presence_of_noise,8006977} consider CQGT with adversarially perturbed measurements. 
%, which assumes the perturbation is adversarial and aims to attain certain detection performance exactly, that is, . 
It has been shown in \cite{8006977} that, for $\delta,\kappa\in (0,1)$, when the perturbation is at most the order of $\Theta(n^\delta)$ and the goal is to detect the defective items within Hamming distance at most the order of $\Theta(n^\kappa)$, there is a sharp phase transition in the fundamental limit: for $0 < 2\delta \leq \kappa < 1$, the optimal \emph{non-adaptive} pooling complexity is $\Theta(\frac{n}{\log n})$, and for $0 < \kappa < 2\delta < 1$, it is $\omega(n^p)\ \forall\,p\in\mbb{N}$. This sharpened results in previous works related to data privacy \cite{Dinur2003RevealingIW,dwork2008new} and pointed out that the relevant regime is $0 < 2\delta \leq \kappa < 1$. 
However, for this more interesting regime where the perturbation is sufficiently small to allow sublinear-in-$n$ pooling complexity, only the existence of optimal measurement plans was shown in \cite{8006977} by a probabilistic argument, while the optimal explicit construction remained open. This is in sharp contrast to the noiseless case, where not only was the optimal pooling complexity characterized in \cite{Erdgs2001ONTP}, but an elegant combinatorial construction of the \emph{non-adaptive} measurement plan was given in \cite{lindstrom_1965,cantor_mills_1966}. Note that deterministic and non-adaptive pooling methods are more favorable in many practical applications: deterministic constructions leverage specific structures and usually lead to low-complexity decoding algorithms; meanwhile, non-adaptive pooling algorithms, in contrast to adaptive ones, allow parallelization of measurements and greatly reduce the overall measuring time.

In this work, significant progress is made in mending the aforementioned gap for non-adaptive CQGT with adversarially perturbed measurements in the regime $0 < 2\delta \leq \kappa < 1$. Our first piece of contribution is sharpening the characterization of the information theoretic limit by refined analyses in the achievability and the converse part. We characterize the relationship between $(\kappa,\delta)$ and the leading coefficient of the optimal non-adaptive pooling complexity which turns out to be $\frac{1}{1-2\delta}\frac{n}{\log n}$ to within a constant factor not depending on $(\kappa,\delta)$. 
We further investigate the sparse CQGT (SCQGT) problem, that is, the original CQGT problem with an additional condition that the number of defective items is not greater than a threshold that we term the \emph{sparsity} level. When the sparsity level is $\Theta(n^\lambda)$ where $0<2\delta \le \kappa < \lambda <1$, the optimal pooling complexity is also characterized to within a constant factor not depending on $(\kappa,\delta,\lambda)$. For these information theoretic results, achievability is proved via a probabilistic argument, and the converse proof extends that of Erd\"{o}s and R\'{e}nyi \cite{Erdgs2001ONTP}. To obtain the tight characterization of the leading coefficient, in the achievability part, we divide all pairs of data vectors (each data vector is a vector of bits that encodes a possible set of locations of defectives) into three different cases and carefully bound the probability that a random matrix cannot distinguish these data pairs in each of the three cases.

%We then note that while a probabilistic argument suffices to prove the existence of asymptotically optimal CQGT measurement plans and establishes the achievability part, an explicit structured construction is preferred in many realistic applications, as the underlying structures often lead to low-complexity decoding algorithm. Our second piece of contribution is an explicit construction of a pooling-complexity optimal deterministic non-adaptive measurement plan, together with a low-complexity decoding algorithm.
%We then note that 
While a probabilistic argument suffices to prove the existence of asymptotically optimal CQGT  measurement plans and establishes the achievability part, an explicit structured construction is preferred in many practical applications, as mentioned previously. 
%, as the underlying structures often lead to low-complexity decoding algorithms. Among all explicit constructions, deterministic and non-adaptive querying algorithms are more favorable in many practical applications. In contrast to a random one, deterministic construction has a fixed structure and usually leads to low-complexity decoding algorithms. Non-adaptive construction, in contrast to adaptive one, can make all tests parallelly in one stage. This parallelization dramatically reduces the overall measuring time since each test takes time in many practical applications. 
Our second piece of contribution is the explicit construction of deterministic and non-adaptive measurement plans for the CQGT problem, together with low-complexity decoding algorithms. 
%As for non-adaptive CQGT algorithms, 
Our construction is pooling-complexity optimal up to a constant factor not depending on $\kappa$ and $\delta$. We first provide an explicit construction of a non-adaptive measurement plan with pooling complexity being $\frac{1}{\kappa-2\delta}\frac{n}{\log_{2} n}$ (suboptimal leading coefficient in terms of $\kappa$ and $\delta$) to within a constant factor. To obtain an optimal leading coefficient $\frac{1}{1-2\delta}\frac{n}{\log_{2} n}$, it turns out that the key is to differentiate data vectors that are within sublinear-in-$n$ Hamming distance, which is equivalent to solving an SCQGT problem. 
We then give an explicit construction of non-adaptive SCQGT algorithms with pooling complexity $o(n^{\lambda+\rho})$ for all $\rho > 0$ and combine it with the aforementioned suboptimal CQGT measurement plan to complete our construction. The pooling complexity of the resulting non-adaptive CQGT measurement plan is $\frac{1}{1-2\delta}\frac{n}{\log_{2} n}$ to within a constant factor (optimal in the leading coefficient).
%a CQGT problem with a sparsity constraint on the to-be-detected data vector.}  

The explicit construction of SCQGT algorithms is highly non-trivial. Note that unlike adaptive pooling and randomized pooling methods, there are no general principles to construct non-adaptive deterministic pooling algorithms in a variety of related problems. For the adaptive case, one can follow the principle of generalized binary splitting algorithms \cite{generalized-binary-splitting}, as in noisy adaptive compressed sensing \cite{CS-adaptive}, noisy adaptive group testing \cite{GT-adaptive}, and noisy quantitative group testing \cite{QGT-adaptive}, to sequentially locate defective items. For constructing random and non-adaptive designs, there also exist general principles, such as (1) sparse-graph-code-based constructions in noisy compressed sensing \cite{CS-sparse-graph-code}, noisy group testing \cite{GT-sparse-graph-code}, and quantitative group testing \cite{QGT-sparse-graph-code}, or (2)  leveraging randomness to equivalently perform adaptive search non-adaptively \cite{GT-adaptive-to-non-adaptive-by-randomness}. Unfortunately, these principles cannot be extended to construct deterministic and non-adaptive pooling algorithms. Existing constructions of deterministic and non-adaptive pooling/sensing methods in group testing \cite{GT-non-adaptive-non-random} and compressed sensing \cite{CS-deterministic-survey} are problem-specific. They all take an algebraic approach (typically taken from the theory of error-correcting codes) and are hence difficult to extend to different noise models. Once the types of measurement and noise models change, it becomes a brand new combinatorial problem, and one needs to solve it all over again.

The main idea behind our solution to the SCQGT problem, called an \emph{embedding approach} hereafter, breaks down the whole construction into two steps. In the first step, we consider the same problem but without a sparsity constraint, that is, the noisy CQGT problem. We aim to construct a close-to-optimal pooling matrix $\mb{M}$ (a pooling matrix is a $\{0,1\}$-matrix that encodes the measurement plan) for the noisy CQGT problem in this step. Note that $\mb{M}$ is different from the suboptimal construction for the overall noisy CQGT problem mentioned earlier. Because the constructed pooling matrix $\mb{M}$ needs to be ``robust'' in the sense that it remains a good pooling matrix even some of its elements are changed so that it can be combined with the non-perfect ``embedding'' in the second step.  In the second step, we aim to embed many such $\mb{M}$'s constructed in the first step into a big overall pooling matrix $\mb{M}_{s}$ for the noisy SCQGT problem, so that whatever the support set (the location of defectives) is, the matrix formed by the corresponding columns of $\mb{M}_{s}$ always contain a submatrix covering most of the columns of $\mb{M}$. Consequently, $\mb{M}_{s}$ is a good pooling matrix for the noisy SCQGT problem. Note that this embedding is non-perfect because it can only guarantee to cover most instead of all of the columns of $\mb{M}$. Therefore the embedded matrix $\mb{M}$ needs to be ``robust'' as mentioned earlier.  
%Moreover, it turns out that we do not need an optimal construction. Instead, a close-to-optimal construction suffices. 

To be more specific, in our construction we leverages a specific combinatorial structure, \emph{lossless unbalanced bipartite expanders} \cite{Ta-ShmaUmans_07, expander_construction}, to achieve the goal of embedding. While the existence of unbalanced bipartite expanders with the optimal parameters can be proved via probabilistic methods, the explicit construction that achieves the optimal parameters remain open to the best of our knowledge. Fortunately, the near-optimal construction by Guruswami \textit{et al.} \cite{expander_construction} suffices to meet our need.  As a result, we provide an explicit construction of a non-adaptive measurement plan for the SCQGT problem with suboptimal pooling complexity ($o(n^{\lambda+\rho})$ for all $\rho > 0$). It is worth noting that our embedding approach is not limited to the noisy SCQGT problem. It suggests a general principle to construct deterministic and non-adaptive pooling algorithms for problems with noise and a sparse structure. We have a detailed discussion in Section~\ref{sec:conclusion}.

Part of the materials of this paper was published in the conference version \cite{LiWang_21}. In this journal version, the complete solution to the explicit construction of the deterministic non-adaptive CQGT measurement plan is given, which is the novel part not presented in \cite{LiWang_21}.

\subsection*{Related Works}
There are several closely related works about CQGT with adversarially perturbed measurements\cite{presence_of_noise,ChangWeldon_79,Wilson_88,ChengKamoi_06}. However, their noise model are all quite different from ours. In \cite{presence_of_noise}, there are three possible outcomes: the correct sum, an erroneous outcome with an arbitrary value, and an erasure symbol "?". When the total number of erroneous (or erasure) outcomes is assumed to be at most a fraction of the total number of measurements,
%of the order of $O(n^{1-\epsilon})$ for some $\epsilon \in (0,1)$, 
which can be viewed as a \emph{$\ell_0$-norm} constraint on the perturbation vector, 
%, while the noise model in our work can be viewed as an \emph{infinity-norm constraint}. 
the optimal non-adaptive pooling complexity is characterized to within a constant factor. Another line of related works pertain to the binary multiple-access adder channel \cite{ChangWeldon_79,Wilson_88,ChengKamoi_06}, where the perturbation vector is constrained in the \emph{$\ell_1$-norm}.  
In contrast, the noise model in our work constrains the perturbation vector in the \emph{$\ell_\infty$-norm}, which makes perfect detection impossible, while in the related works mentioned above, only the perfect-detection criterion is considered. 
%a non-adaptive algorithm with $O(\frac{n}{\epsilon \log n})$ pooling complexity is proposed in \cite{presence_of_noise}, but there is no matching lower bound. 

Other related group testing problems are discussed as follows. Threshold group testing (TGT) deals with approximate recovery under adversarial noise with binary response. Their pooling responses can be directly reduced from ours by passing through an additional three-level quantizer with a lower threshold $l$ and an upper threshold $u$.
\begin{align*}
f(x)=\begin{Bmatrix} 
   0 ,& x\leq l\\
   \text{arbitrary 0,1}, & l<x<u\\
   1, & u\leq x\\
   \end{Bmatrix}
\end{align*}
Consequently, our lower bound holds in their setting, and their algorithms can always be transformed to our setting. However, existing literature about TGT are all focused on either the case with a zero-gap $g=0$ ($g:=u-l-1$) \cite{TGT-gap0-2020,TGT-gap0-2019} or approximately recover data set to within $\Theta(g)$ errors \cite{TGT-withinG-1,TGT-withgap-2,TGT-original}, which corresponds to a more stringent recovery criteria than ours. Group testing with adversarial flipped noise \cite{GT-advflip} has an incomparable noise model with ours. It uses $m=O(d\log(n))$ measurements to recover any $d$-sparse binary vectors within $O(d)$ decoding errors with up to $\delta m$ false positive and $O(m/d)$ false negative in pooling responses, for any constant $\delta<1$. Semiquantitative group testing (SQGT) \cite{SQGT-original,SQGT-2021-expander} consider noiseless QGT with an additional integer-valued quantizer at the output of the measurements. It can always achieve perfect recovery. It is worth mentioning that in \cite{SQGT-2021-expander}, their construction also uses lossless unbalanced bipartite expanders, but with a different method and high-level intuition from ours.

The rest of the paper is organized as follows. In Section~\ref{sec:formulation}, we introduce the noisy CQGT and SCQGT problems and provide results of fundamental limits. In Section~\ref{sec:algo}, we explicitly construct pooling designs and corresponding decoding algorithms for noisy CQGT and SCQGT problems. In Section~\ref{sec:fundamental_proof}, we prove theorems of fundamental limits stated in Section~\ref{sec:formulation}. In Section~\ref{sec:conclusion}, we conclude our work, provide several future directions, and discuss about a general principle extended from our embedding method to construct deterministic and non-adaptive pooling algorithms for problems with noise and a sparse structure.

%\section{Problem Formulation and Fundamental Limits}\label{sec:formulation_and_infolimit}
\section{Quantitative Group Testing Problem and Its Fundamental Limits}\label{sec:formulation}
%In this paper, $k_{n}, \delta_{n}, l_{n}$ means that $k, \delta, l$ are function of $n$. And in most of the theorem and statement, we deal with the special case $k_{n} = n^{k}, \delta_{n} = n^{\delta}, l_{n} = n^{l}$, where $k,\delta,l\in \lp 0,1\rp$.

In this section, let us first define the combinatorial quantitative group testing (CQGT) problem and other related notions, and then provide the characterization of its fundamental limits.
\subsection{Problem formulation}
The following are the key elements of a CQGT problem.
\begin{enumerate}
\item 
Data: for each item indexed by $j=1,...,n$, we use $x_j\in\{0,1\}$ to denote whether or not the $j$-th item is defective. Hence, the $n$-by-$1$ \emph{data vector} 
\[
\bm{x} := \begin{bmatrix}
x_1\\ x_2\\ \vdots\\ x_n
\end{bmatrix}
\]
is the target to be reconstructed from the noisy measurements. 
%      the possible value $x_1,...,x_n\in \{0,1\}$, we collect $x_1,...,x_n$ into a single $n$ by $1$ binary status vector $\bm{X}$.

\item
Counting measurements: the pool of items in the $i$-th counting measurement can be represented by an $1$-by-$n$ \emph{pooling vector} $\bm{r_{i}}\in \{0,1\}^{1\times n}$, and the outcome of the counting measurement is $\bm{r}_{i}\bm{x}$. For a non-adaptive pooling algorithm, the measurement plan can be concisely represented by an $n$-by-$s$ \emph{pooling matrix} 
\[
\mb{R} := \begin{bmatrix}
\bm{r}_1\\ 
\bm{r}_2\\ \vdots\\ \bm{r}_s
\end{bmatrix}
\] with its $i$-th row being the $i$-th pooling vector $\bm{r}_i$. Here $s$ denotes the number of measurements, termed \emph{pooling complexity}.
%      and the corresponding response $y_{i} = \bm{r}_{i}\bm{X} + n_i$,  where $n_i$ is a real number represent the noise. For a non-adaptive pooling algorithm, we can represent it by a pooling matrix $\mb{Q}$ with $i$-th row corresponding to $i$-th measurement, and collect $\{y_i\},\{n_i\}$ into response vector $\bm{Y}$, noise vector $\bm{N}$. So, $\bm{Y}=\mb{Q}\bm{X}+\bm{N}$. 

\item
Perturbed outcomes: the outcome of the $i$-th measurement is $y_{i} = \bm{r}_{i}\bm{x} + n_i$, where $n_i\in [-d_n,d_n]$ denotes the bounded additive perturbation in the $i$-th measurement. The $s$ outcomes of the measurements can be concisely written as an $s$-by-$1$ vector 
\[
\bm{y} := \begin{bmatrix}
y_1\\ y_2\\ \vdots\\ y_s
\end{bmatrix}
= \mb{R}\bm{x} + \bm{n},
\]
where $\bm{n}$ is the \emph{perturbation vector} with $\lnorm \bm{n}\rnorm _{\infty}\leq d_n$.
%    For each measurement $\bm{q_i}$, the corresponding noise $n_i$ can be arbitrary real number with absolute value smaller than $\delta_n$, hence $\lnorm \bm{N}\rnorm _{\infty}\leq d_n$.  

\item
Detection: for any data vector $\bm{x}\in\{0,1\}^{n\times 1}$, the estimate generated by the detection algorithm (denoted by $\hat{\bm{x}}$) should be close to $\bm{x}$. In particular, the Hamming distance between $\hat{\bm{x}}$ and $\bm{x}$ should not be greater than $k_n$, that is, $\lVert\hat{\bm{x}}-\bm{x}\rVert_0 \le k_n$.
\end{enumerate}

A pooling matrix $\mb{R}$ is said to solve the above CQGT problem if 
\begin{equation}\label{eq:detect_criterion}
\begin{array}{l}
\lVert \mb{R}\bm{x} - \mb{R}\bm{x}'\rVert_\infty > 2d_n,\quad \forall\,\bm{x},\bm{x}'\in\{0,1\}^{n\times 1}\ \text{with}\ \lVert\bm{x} - \bm{x}'\rVert_0 > k_n.
\end{array}
\end{equation}
Let us introduce the following definition.
\begin{definition}\label{def:CQGT}
%\textsf{$(n,k_n,d_n)$-CQGT} 
\CQGT{n,k_n,d_n} denotes the combinatorial quantitative group testing problem defined above. 
If a pooling matrix $\mb{R}$ is a solution to \CQGT{n,k_n,d_n}, it is called an $(n,k_n,d_n)$-detecting matrix. 
$s_{\textsf{CQGT}}^*(n,k_n,d_n)$ denotes the smallest possible pooling complexity among all non-adaptive pooling algorithms, that is, it is the smallest height of $(n,k_n,d_n)$-detecting matrices.
\end{definition}

%Our goal is to characterize $s_{\textsf{CQGT}}^*(n,k_n,d_n)$ in the asymptotic regime where $d_n = \Theta(n^\delta)$ and $k_n = \Theta(n^\kappa)$, for $\delta,\kappa\in(0,1)$. This gives the following definition:

%One can already seen that our criteria is combinatorial criteria, which is stronger than w.h.p criteria in \cite{8006977}. So we can see that if $\delta_{n} = n^{\delta}, k_{n} = n^{k}, 2\delta>k$, then $s^{*}=\Omega\lp n^p\rp , \forall p\in N$ immediately. Hence, in this work, we focus on the case $\delta_{n} = n^{\delta}, k_{n} = n^{k}, 0<2\delta\leq k<1$.

Throughout our development, it turns out that CQGT with an additional sparsity constraint, which we call \emph{sparse combinatorial group testing} (SCQGT), can be and should be explored simultaneously, as it serves as part of our explicit construction of non-adaptive CQGT measurement plans. 
%The optimal pooling complexity for non-adaptive SCQGT is also characterized, along with efficient adaptive algorithms. 
Let us introduce the following definition to better refer to this problem.

\begin{definition}\label{def:SCQGT}
%$\lp n,k_{n},d_{n},l_{n}\rp$-SCQGT 
\SCQGT{n,k_n,d_n,l_n} denotes the combinatorial quantitative group testing problem \CQGT{n,k_n,d_n} with the additional sparsity assumption on the data vector $\bm{x}$, that is, $\lVert \bm{x}\rVert _{0}\leq l_{n}$. 
If a pooling matrix $\mb{R}$ is a solution to \SCQGT{n,k_n,d_n,l_n}, with a slight abuse of notation, it is called an $(n,k_n,d_n,l_n)$-detecting matrix. 
$s_{\textsf{SCQGT}}^*(n,k_n,d_n,l_n)$ denotes the smallest possible pooling complexity among all non-adaptive pooling algorithms, that is, it is the smallest height of $(n,k_n,d_n,l_n)$-detecting matrices.
\end{definition}

%\begin{definition}
%For non-adaptive algorithm, if the pooling matrix $\mb{Q}$ satisfy our recovery criteria of $\lp n,k_{n},\delta_{n}\rp $-CQGT, then we say $\mb{Q}$ is a $\lp n,k_{n},\delta_{n}\rp $-detecting matrix. If the pooling matrix $\mb{Q}$ satisfy our recovery criteria of $\lp n,k_{n},\delta_{n},l_{n}\rp $-SCQGT, then we say $\mb{Q}$ is a $\lp n,k_{n},\delta_{n},l_{n}\rp $-detecting matrix.
%\end{definition}

%\subsection{Fundamental limits for the non-adaptive case}
\subsection{Fundamental limits on the pooling complexity}
Our first piece of contribution is the characterization of the optimal non-adaptive pooling complexity for \CQGT{n,n^\kappa,n^\delta}, $0<2\delta \le \kappa <1$. The characterization is tight to within a constant factor that is independent of $(n, \kappa, \delta)$, as stated in the following theorem.

\begin{theorem}\label{QGT coro}
     For $0<2\delta \le \kappa <1$, 
\[
s_{\textsf{CQGT}}^*(n,n^{\kappa},n^{\delta}) = \frac{1}{1-2\delta}\frac{n}{\log n}
\]
up to a constant factor that is independent of $(n, \kappa, \delta)$.
%     , and it's independent of $\kappa$. 
\end{theorem}
\begin{IEEEproof}
The proof comprises two parts: achievability and converse, established in the lemmas below. 

%For the achievable part, we use random pooling to show that.
\begin{lemma}[CQGT Achievability]\label{non_constructive_proof}
For $0<2\delta \le \kappa <1$,
\[
\limsup_{n\ra\infty} \frac{s_{\textsf{CQGT}}^*(n,n^{\kappa},n^{\delta})}{n/\log n}\le \frac{8}{1-2\delta}\ .
\]
In words, there exists a sequence of $(n,n^\kappa,n^\delta)$-detecting matrices with pooling complexity not greater than
%  deterministic non-adaptive pooling algorithm for \CQGT{n,n^\kappa,n^\delta}, with pooling complexity asymptotically goes to 
$\frac{8}{1-2\delta}\frac{n}{\log n}$ as $n\ra\infty$.
\end{lemma}

%We first derived $s_{\textsf{CQGT}}^*(n,n^{\kappa},n^{\delta})$. For the converse part, inspired by the method in\cite{Erdgs2001ONTP}, we prove 
\begin{lemma}[CQGT Converse]\label{lower_bound} For $0<2\delta \le \kappa <1$, 
\[
s_{\textsf{CQGT}}^*(n,n^{\kappa},n^{\delta})\geq \frac{2}{1-2\delta}\frac{n}{\log n}.
\]
\end{lemma}

The two lemmas complete the proof of the theorem. 
The proof of achievability (Lemma~\ref{non_constructive_proof}) is in Section~\ref{subsec:pf_direct}, which uses a probabilistic argument to prove the existence of good pooling matrices. Converse (Lemma~\ref{lower_bound}) is proved in Section~\ref{subsec:pf_converse}, which is based on extending a counting argument with its root in \cite{Erdgs2001ONTP}.
\end{IEEEproof}

%We prove Theorem \ref{lower_bound} by a counting argument. In order to satisfy detection criteria, any $\lp n,n^{\kappa},n^{\delta}\rp $-detecting matrix $\mb{Q}$ must have, $\forall \lnorm\bm{x}-\bm{\hat{x}}\rnorm _{0}\geq n^{\kappa}, \lnorm\mb{Q}\bm{x}-\mb{Q}\bm{\hat{x}}\rnorm _{\infty}\geq n^{
%\delta}$. Hence we can lower bound the pooling complexity by the packing number of $\{\mb{Q}\bm{x}\lvert\bm{x}\in\{0,1\}^{n}\}$ with radius $n^{\delta}$ under infinity norm.

%Combine Theorem $\ref{lower_bound}$ and Theorem $\ref{non_constructive_proof}$.We get

%Later in the code construction(\ref{Explicit measurement plans}), instead of constructing an  optimal non-adaptive pooling algorithm achieving $s_{\textsf{CQGT}}^*(n,n^{\kappa},n^{\delta})$, we only construct an adaptive algorithm achieving $s_{\textsf{CQGT}}^*(n,n^{\kappa},n^{\delta})$. Hence the optimal non-adaptive construction is still open. For the construction of this adaptive algorithm, we will divide original problem($\lp n,n^{\kappa},n^{\delta}\rp $-CQGT) into two sub problem ( $\lp n,n^{\lambda},n^{\delta}\rp $-CQGT and $\lp n,n^{\kappa},n^{\delta},n^{\lambda}\rp $-SCQGT). Hence, we going to derive $s_{\textsf{SCQGT}}^*(n,n^{\kappa},n^{\delta},n^{\lambda})$. First we show the lower bound by similar approach in Theorem \ref{lower_bound} .

It is interesting to note that the leading coefficient does not depend on the order of the detection criterion $\kappa$. In other words, as long as partial detection to within $n^\kappa$ successfully detection items is allowed, the number of pools to be measured only depend on the strength of the adversarial perturbation $n^\delta$, where $\delta \le \kappa/2$. 
%\commentihw{Add some high-level explanation why}

When the defective items are sparsely populated in the data set, the number of pools needed to be measured should be smaller. The following theorem characterizes the optimal non-adaptive pooling complexity for \SCQGT{n,n^\kappa,n^\delta,n^\lambda} when $0<2\delta \le \kappa<\lambda <1$.

\begin{theorem}%[fundamental limit of $\lp n,n^{\kappa},n^{\delta},n^{\lambda}\rp $-SCQGT]
\label{thm:SCQGT_limit}
For $0<2\delta \le \kappa < \lambda <1$, % and $0<\lambda<1$,
\[
s_{\textsf{SCQGT}}^*(n,n^{\kappa},n^{\delta},n^{\lambda})=
\begin{cases}
                \frac{1-\lambda}{\lambda-2\delta}n^\lambda,  &2\delta<\kappa   \\[1ex]
                \frac{1-\lambda}{\lambda-2\delta}n^{\lambda}\log n,  &2\delta=\kappa
\end{cases}
\]
up to a constant factor that is independent of $(n,\kappa,\delta,\lambda)$. 
\end{theorem}
\begin{IEEEproof}
Similar to the proof of Theorem~\ref{QGT coro}, the following two lemmas correspond to achievability and converse respectively, and their combination completes the proof.

\begin{lemma}[SCQGT Achievability]\label{sparse_achievability}
For $0\!<\! 2\delta \!\le\! \kappa \!<\! \lambda <\!1$, % and $0<\lambda<1$,
\begin{align*}\textstyle
\limsup_{n\ra\infty} \frac{s_{\textsf{SCQGT}}^*(n,n^{\kappa},n^{\delta},n^{\lambda})}{n^\lambda}
&\textstyle\le \frac{4(1-\lambda)}{\lambda-2\delta},  && 2\delta<\kappa\\
\textstyle\limsup_{n\ra\infty} \frac{s_{\textsf{SCQGT}}^*(n,n^{\kappa},n^{\delta},n^{\lambda})}{n^{\lambda}\log n}
&\textstyle\le \frac{4(1-\lambda)}{\lambda-2\delta},  && 2\delta=\kappa
%\begin{cases}
%                \frac{4(1-\lambda)}{\lambda-2\delta}n^\lambda,  &2\delta<\kappa   \\[1ex]
%                \frac{4(1-\lambda)}{\lambda-2\delta}n^{\lambda}\log n,  &2\delta=\kappa
%\end{cases}\ .
%\frac{8}{1-2\delta}\frac{n}{\log n},
\end{align*}
%
%    There exists a deterministic non-adaptive pooling algorithm for $\lp n,n^{\kappa},n^{\delta},n^{\lambda}\rp $-SCQGT with pooling complexity asymptotically goes to
%    
%    \begin{equation*}
%        s = 
%        \lbp
%            \begin{array}{cc}
%                \frac{4\lp 1-\lambda\rp n^\lambda}{\lambda-2\delta}  & ,2\delta<\kappa   \\
%                \frac{4\lp 1-\lambda\rp n^{\lambda}\log\lp n\rp }{\lambda-2\delta} & ,2\delta=\kappa
%            \end{array}
%        \rbp
%    \end{equation*}
\end{lemma}

\begin{lemma}[SCQGT Converse]\label{sparse_converse}
For $0<2\delta \le \kappa < \lambda <1$, % and $0<\lambda<1$,
\[
s_{\textsf{SCQGT}}^*(n,n^{\kappa},n^{\delta},n^{\lambda}) \ge
\begin{cases}
                \frac{2(1-\lambda)}{\lambda-2\delta}n^\lambda,  &2\delta<\kappa   \\[1ex]
                \frac{2(1-\lambda)}{\lambda-2\delta}n^{\lambda}\log n,  &2\delta=\kappa
\end{cases}
\]
\end{lemma}

Proofs of the above two lemmas are similar to those of Lemma~\ref{non_constructive_proof} and \ref{lower_bound} and hence are appended after the proof of Lemma~\ref{non_constructive_proof} and \ref{lower_bound}.
% of the extended version.
%Next we show achievability by random pooling similar to Theorem \ref{non_constructive_proof}.
\end{IEEEproof}
%Combine lemma $\ref{sparse_converse}$ and lemma $\ref{sparse_achievability}$. We get 

%\section{Main Results}\label{sec:results}

\section{Explicit Construction of an Asymptotically Optimal Non-Adaptive Pooling}\label{sec:algo}

{In this section, first we give a basic construction of a non-adaptive measurement plan for \CQGT{n,n^{\kappa},n^{\delta}} that can only effectively distinguishing those data pairs with linear-in-n Hamming distances. If we also apply it to data pairs with sublinear-in-n Hamming distances, then the overall pooling complexity would has the optimal order in $n$ but a suboptimal leading coefficient in terms of $\kappa, \delta$ in Section~\ref{Basic_construction}. 
To achieve a better leading coefficient, a non-adaptive pooling algorithm for \textsf{SCQGT} is developed for data pairs with sublinear-in-n Hamming distances via the embedding method in Section~\ref{adaptive_embedding}. This non-adaptive pooling algorithm is then combined with the basic non-adaptive measurement plan in Section~\ref{Basic_construction} to give an explicit construction of a non-adaptive pooling algorithm with optimal leading coefficient of the pooling complexity in Section~\ref{subsec:final_form}, along with an efficient (nearly-linear-in-$n$) decoding algorithm. All results about decoding algorithms are elaborated in Section~\ref{subsec:decoding}.} 
%Then, we propose a construction that can achieve the optimal leading coefficient when efficient algorithms for \SCQGT{n,n^{\kappa},n^{\delta},n^{\lambda}} are available. 
%To this end, the design problem boils down to finding good measurement plans for \SCQGT{n,n^{\kappa},n^{\delta},n^{\lambda}}. We give the explicit construction of a sufficiently good adaptive measurement plan for \SCQGT{n,n^{\kappa},n^{\delta},n^{\lambda}} in Section~\ref{adaptive_embedding}. While the explicit construction of an efficient non-adaptive measurement plan for \SCQGT{n,n^{\kappa},n^{\delta},n^{\lambda}} is still missing, we provide reduction of this problem to a simpler combinatorial problem detailed in Section~\ref{reduction_SCQGT}.   

\subsection{A basic construction for distinguishing data pairs with linear-in-n Hamming distances}\label{Basic_construction}
%We going to explicit code construction for \CQGT{n,n^{\kappa},n^{\delta}}. First we construct a simple non-adaptive pooling algorithm. The key ingredient of this construction is the structure of Hadamard matrix. 

%In the following construction, we use two different matrices as our building blocks, sometimes these matrix do not have the size we want. For example, in the following construction, we need a Sylvester's type Hadamard matrix with size $n^{1-\frac{\epsilon}{2}}$, but the size of Sylvester's type Hadamard matrix must be $2^i,i\in N$, and $n^{1-\frac{\epsilon}{2}}$ might not equal to any $2^i,i\in N$. To fix this problem, each time when we don't have building block with correct size, we simply find the building block matrix with minimum size that greater or equal than our desire. So the result $\lp n,n^{\kappa},n^{\delta}\rp $-detecting matrix has width slightly bigger than $n$. We then simply delete some column of result matrix until width equal to $n$, this matrix will also be a $\lp n,n^{\kappa},n^{\delta}\rp $-detecting matrix(by the definition of $\lp n,n^{\kappa},n^{\delta}\rp $-detecting matrix). And one can easily check that, because of additional manipulation mentioned above, pooling complexity would multiply some additional constant independent of $k, d$. Hence the optimality in terms $k, d$ remains.

The basic construction of the non-adaptive CQGT measurement plan is given below. 
%, which leverage the structure and properties of the Hadamard matrix 
Some necessary notations are set up first. Let $\epsilon = \kappa-2\delta > 0$. 
Let $\slceil n^{\epsilon/2}\srceil$ denote the smallest possible width of the detecting matrix for the noiseless coin weighing problem mentioned in Section 4 of \cite{cantor_mills_1966} that is not smaller than $n^{\epsilon/2}$, and $\mb{M}_{\slceil n^{\epsilon/2}\srceil}$ be the corresponding detecting matrix. 
Let $\hlceil n^{1-\epsilon/2}\hrceil$ denote the smallest possible size of the Hadamard matrix (Sylvester's construction) that is not smaller than $n^{1-\epsilon/2}$, and $\mb{H}_{\hlceil n^{1-\epsilon/2}\hrceil}$ be the corresponding Hadamard matrix. 
Let $\bar{n} = \slceil n^{\epsilon/2}\srceil\,\hlceil n^{1-\epsilon/2}\hrceil$ and let 
\[
\mb{Q}_{\bar{n}}^{0} = \mb{H}_{\hlceil n^{1-\epsilon/2}\hrceil} \otimes \mb{M}_{\slceil n^{\epsilon/2}\srceil}
\] 
the Kronecker product of the two matrices. 

We are ready to give our basic construction. According to the above setup, entries of $\mb{Q}_{\bar{n}}^{0}$ take value in $\{0,\pm1\}$. 
Let us find two $\{0,1\}$-matrices $\mb{Q}_{\bar{n}}^{1}$ and $\mb{Q}_{\bar{n}}^{2}$ such that $\mb{Q}_{\bar{n}}^{1}-\mb{Q}_{\bar{n}}^{2} = \mb{Q}_{\bar{n}}^{0}$, concatenate them vertically into a new matrix $\mb{Q}_{\bar{n}}^{3}$, and delete the last $\bar{n}-n$ columns of $\mb{Q}_{\bar{n}}^{3}$ to get $\mb{Q}$. The width of the matrix $\mb{Q}$ becomes $n$, and $\mb{Q}$ stands for the measurement matrix that we would like to construct. 
By leveraging the structure of the Hadamard matrix along with the detecting capability of $\mb{M}$, we show that $\mb{Q}$ is a detecting matrix for CQGT with it pooling complexity summarized in the following theorem. 
%guarantees summarized in following theorem, the proof of which is detailed in Appendix~\ref{pf_basic_construction}. 
As for decoding, later in Section~\ref{subsec:decoding}, 
%Appendix~\ref{decode algorithm}, 
we provide a companion two-step decoding algorithm for this non-adaptive measurement plan and show that it has $O(n)$ time complexity.

\begin{theorem}[Basic Construction]\label{basic_construction}
For $n$ sufficiently large, $\mb{Q}$ is a $\lp n,n^{\kappa},n^{\delta}\rp $-detecting matrix with pooling complexity no more than
    $\frac{48}{\kappa-2\delta}\frac{n}{\log_{2}\lp n\rp}$.
%    , and an efficient $O(n)$ decoding algorithm.
\end{theorem}
\begin{IEEEproof}
Let $\epsilon\in \lp0,1\rp$, and $\slceil{n^{\frac{\epsilon}{2}}}\srceil$ denote the smallest width of noiseless detecting matrix corresponding to non-adaptive pooling algorithm mentioned in Section 4 of \cite{cantor_mills_1966} that greater or equal than $n^{\frac{\epsilon}{2}}$. Let $\slceil{n^{1-\frac{\epsilon}{2}}}\srceil$ denote the smallest size of Sylvester's type Hadamard matrix that greater or equal than $n^{1-\frac{\epsilon}{2}}$. Let $\bar{n} = \slceil{n^{1-\frac{\epsilon}{2}}}\srceil\slceil{n^{\frac{\epsilon}{2}}}\srceil$. One can easily get that ${n^{\frac{\epsilon}{2}}}\leq\slceil{n^{\frac{\epsilon}{2}}}\srceil\leq 3{n^{\frac{\epsilon}{2}}}$, and ${n^{1-\frac{\epsilon}{2}}}\leq\slceil{n^{1-\frac{\epsilon}{2}}}\srceil\leq 2{n^{1-\frac{\epsilon}{2}}}$, and consequently, $n\leq\bar{n} \leq 6n$. 
%It worth to mention that since $\mb{M}_{\slceil n^{\frac{\epsilon}{2}}\srceil}$ is a noiseless detecting matrix, 

In order to proof that $\mb{Q}$ is a $\lp n,n^{\kappa},n^{\delta}\rp $-detecting matrix, we need to show that $\forall \bm{a}\neq \bm{b}\in \{0,1\}^{n} \emph{ with } \lnorm \bm{a}-\bm{b}\rnorm _{0}\geq n^{\kappa}$, $\lnorm \mb{Q}\lp \bm{a}-\bm{b}\rp \rnorm _{\infty}\geq 2n^{\delta}$. Since $\mb{Q}$ is reduced(delete last $\bar{n}-n$ column) from $\mb{Q}_{\bar{n}}^{3}$, and $\mb{Q}_{\bar{n}}^{3}$ is the concatenate of $\mb{Q}_{\bar{n}}^{1},\mb{Q}_{\bar{n}}^{2}$, and $\mb{Q}_{\bar{n}}^{0} = \mb{Q}_{\bar{n}}^{1} - \mb{Q}_{\bar{n}}^{2}$. It suffics to show that for any $\bm{a}\neq \bm{b}\in \{0,1\}^{\bar{n}} \emph{ with } \lnorm \bm{a}-\bm{b}\rnorm _{0}\geq n^{\kappa}$, 
$\lnorm \mb{Q}_{\bar{n}}^{0}\lp \bm{a}-\bm{b}\rp \rnorm _{\infty}\geq 2n^{\delta}$. 

Let $\bm{d}=\bm{a}-\bm{b}=\lb\bm{d}_1,\bm{d}_2,...,\bm{d}_{\slceil n^{1-\frac{\epsilon}{2}}\srceil}\rb$ be the equal length division of some difference vector, where $\bm{a}\neq \bm{b}\in \{0,1\}^{\bar{n}}$. Let $\bm{y} = \lb\bm{y}_1,\bm{y}_2,...,\bm{y}_{\slceil n^{1-\frac{\epsilon}{2}}\srceil}\rb = \mb{Q}_{\bar{n}}^{0}\bm{d}$ be the equal length division of result vector.
Since rows of Hadamard matrix form an orthogonal basis,
\begin{equation*}
    \lnorm \bm{y}\rnorm ^2 = \lnorm \mb{Q}_{\bar{n}}^{0}\bm{d}\rnorm ^2 = \sum_{i=1}^{\slceil n^{1-\frac{\epsilon}{2}}\srceil}\slceil n^{1-\frac{\epsilon}{2}}\srceil\lnorm \mb{M}_{\slceil n^{\frac{\epsilon}{2}}\srceil}\bm{d}_{i}\rnorm ^2
\end{equation*}
Moreover, $\mb{M}_{\slceil n^{\frac{\epsilon}{2}}\srceil}$ is a noiseless detecting matrix. Therefore, $\forall\bm{d}_{i}\neq \bm{0}$, $\mb{M}_{\slceil n^{\frac{\epsilon}{2}}\srceil}\bm{d}_{i}\neq \bm{0}$. Combined with the fact that $\mb{M}_{\slceil n^{\frac{\epsilon}{2}}\srceil}\bm{d}_i$ is an integer vector, we conclude $\lnorm \mb{M}_{\slceil n^{\frac{\epsilon}{2}}\srceil}\bm{d}_i\rnorm ^2\geq 1$. 
But in our setting, $\lnorm \bm{d}\rnorm _0\geq n^{\kappa}$, so, there exists at least $\frac{n^{\kappa}}{\slceil n^{\frac{\epsilon}{2}}\srceil}$ segments $\bm{d}_i\neq \bm{0}$, hence 
\begin{align*}
    \lnorm \bm{y}\rnorm ^2 = \sum_{i=1}^{\slceil n^{1-\frac{\epsilon}{2}}\srceil}\slceil n^{1-\frac{\epsilon}{2}}\srceil\lnorm \mb{M}_{\slceil n^{\frac{\epsilon}{2}}\srceil}\bm{d}_{i}\rnorm ^2
    \geq {\slceil n^{1-\frac{\epsilon}{2}}\srceil}{\frac{n^{\kappa}}{\slceil n^{\frac{\epsilon}{2}}\srceil}}\geq \frac{n^{1+\kappa-\epsilon}}{3}
\end{align*}
finally, remember that the height of $\bm{y}$ equals to the height of $\mb{M}_{\slceil n^{\frac{\epsilon}{2}}\srceil}$ times the height of $\mb{H}_{\slceil n^{1-\frac{\epsilon}{2}}\srceil}$, which is 
\begin{align}
    &\lp\frac{2{\slceil n^{\frac{\epsilon}{2}}\srceil}}{\log_{2}\lp{\slceil n^{\frac{\epsilon}{2}}\srceil}\rp} + O\lp\frac{{\slceil n^{\frac{\epsilon}{2}}\srceil}\log_{2}\lp\log_{2}\lp{\slceil n^{\frac{\epsilon}{2}}\srceil}\rp\rp}{\log_{2}^{2}\lp{\slceil n^{\frac{\epsilon}{2}}\srceil}\rp}\rp\rp  \\
    &\slceil n^{1-\frac{\epsilon}{2}}\srceil \leq \frac{12n}{\log_{2}\lp{ n^{\frac{\epsilon}{2}}}\rp} + O\lp \frac{n\log_{2}\lp\log_{2}\lp n\rp\rp}{\log_{2}^{2}\lp n\rp}\rp \\ &\leq^{(1)} \frac{48n}{\epsilon\log_{2}\lp n\rp} =^{(2)} \frac{48n}{\lp \kappa-2\delta\rp\log_{2}\lp n\rp} = o\lp \frac{n}{3}\rp \label{very_ugly}
\end{align}

When $n$ large enough, inequality ($1$) follows. Equality ($2$) follows from that we choose $\epsilon = \kappa-2\delta$. So we have, 

\begin{equation*}
    \lnorm \bm{y}\rnorm _{\infty}\geq \sqrt{\frac{4\lnorm \bm{y}\rnorm ^{2}}{\frac{n}{3}}} \geq \sqrt{4n^{\lp 1+\kappa-\epsilon\rp -1}} = \sqrt{4n^{\kappa-\epsilon}} =^{(1)} 2n^{\delta}
\end{equation*}

 Equality ($1$) follows from that we choose $\epsilon = \kappa-2\delta$. Hence we can see that, $\forall\bm{a}\neq \bm{b}\in \{0,1\}^{\bar{n}} \emph{ with } \lnorm \bm{a}-\bm{b}\rnorm _{0}\geq n^{\kappa}$, $\lnorm \mb{Q}_{\bar{n}}^{0}\lp \bm{a}-\bm{b}\rp \rnorm _{\infty}\geq 2n^{\delta}$, which implies $\mb{Q}$ is a $\lp n,n^{\kappa},n^{\delta}\rp $-detecting matrix, and by \eqref{very_ugly}, the pooling complexity of this construction is no more than $\frac{48n}{\lp \kappa-2\delta\rp\log_{2}\lp n\rp}$.

\end{IEEEproof}

Compared with the information theoretic limit in Theorem~\ref{QGT coro}, note that for the basic construction, the leading coefficient in the pooling complexity in terms of $\kappa$ and $\delta$ is suboptimal. In other words, if one would like to achieve a better pooling complexity, the error-correcting capability of the basic construction is weakened. Towards improving the leading coefficient in the pooling complexity of the above basic construction, we shall employ 
%in the following we first outline the program of the improved construction. The construction turns out to be the combination of the above basic one and 
a sufficiently good non-adaptive pooling algorithm for \SCQGT{n,n^{\kappa},n^{\delta},n^{\lambda}} to compensate such degradation. The idea goes as follows. 

Choose a $\lambda \in (\kappa,1)$ and take the aforementioned basic construction for \CQGT{n,n^{\lambda},n^{\delta}} with it pooling matrix being $\mb{Q}$.
Suppose we are also given a non-adaptive measurement plan for \SCQGT{n,n^{\kappa},n^{\delta},n^{\lambda}} with it pooling matrix being $\mb{P}$. 
Then, we concatenate $\mb{Q}$ and $\mb{P}$ vertically to get a taller pooling matrix
\begin{equation}\label{decompose}
    \mb{R} = \begin{bmatrix}\mb{Q} \\ \mb{P} \end{bmatrix}
\end{equation}
By definition of the detecting matrices, 
%Since $\lVert\mb{M}x\rVert_{\infty}\geq n^{\delta}$ for all $\bm{x}\in\{0,\pm1\}^{n}$ and $\lVert x\rVert_{0}\geq n^{\lambda}$. 
\begin{align*}
\lVert\mb{Q}\bm{x}\rVert_{\infty} &\geq 2n^{\delta},\quad \forall\,\bm{x}\in\{0,\pm1\}^{n}\text{ with }\lVert \bm{x}\rVert_{0}\geq n^{\lambda},\\
\lVert\mb{P}\bm{x}\rVert_{\infty} &\geq 2n^{\delta},\quad \forall\, \bm{x}\in\{0,\pm1\}^{n}\text{ with }2n^{\lambda}\geq\lVert \bm{x}\rVert_{0}\geq n^{\kappa}.
\end{align*}
%We use the same divide and conquer as in section \ref{adaptive_embedding}, the only difference is that now the building block, $\lp n,n^{\kappa},n^{\delta},n^{\lambda}\rp $ sparse pooling algorithm need to be non-adaptive. In other words, let $\mb{M}$ be the pooling matrix for basic code, and $\mb{Q}$ be the pooling matrix for $\lp n,n^{\kappa},n^{\delta},n^{\lambda}\rp $ sparse code. Then $\begin{bmatrix}\mb{M}\\\mb{Q}\end{bmatrix}$ would be the pooling matrix of a \CQGT{n,n^{\kappa},n^{\delta}} algorithm, because by the definition of basic code, $\mb{M}$ has the property that 
%\begin{align*}
%    \forall \bm{x}\in\{0,1,-1\}^{n}, \lVert x\rVert_{0}\geq n^{\lambda}, \lVert\mb{M}x\rVert_{\infty}\geq n^{\delta}
%\end{align*}
%
%by the definition of \SCQGT{n,n^{\kappa},n^{\delta},n^{\lambda}}, $\mb{Q}$ has the property that
%\begin{align*}
%    \forall \bm{x}\in\{0,1,-1\}^{n}, 2n^{\lambda}\geq\lVert x\rVert_{0}\geq n^{\kappa}, \lVert\mb{Q}x\rVert_{\infty}\geq n^{\delta}
%\end{align*}
%
Hence, $\lVert\mb{R}\bm{x}\rVert_{\infty} 
= \max\{\lVert\mb{Q}\bm{x}\rVert_{\infty},\lVert\mb{P}\bm{x}\rVert_{\infty}\} \geq 2n^{\delta}
$ for all $\bm{x}\in\{0,\pm1\}^{n}\text{ with }\lVert \bm{x}\rVert_{0}\geq n^{\kappa}$. This suggests that $\mb{R}$, the vertical concatenation of $\mb{Q}$ and $\mb{P}$, is a detecting matrix for \CQGT{n,n^{\kappa},n^{\delta}}. 
Its pooling complexity is the sum of those of $\mb{P}$ and $\mb{Q}$. 
Since the pooling complexity of $\mb{Q}$ is asymptotically no more than $\frac{48}{\lambda-2\delta}\frac{n}{\log_{2}n}$, as long as that of $\mb{P}$ is $o(\frac{n}{\log n})$, the overall pooling complexity can be made $\frac{48}{1-2\delta}\frac{n}{\log_{2}n}$ to within a constant factor by setting $\lambda \ra 1$.

\subsection{Explicit construction of the pooling matrix for distinguishing data pairs with sublinear-in-n Hamming distances}\label{adaptive_embedding}
With the discussion in Section~\ref{Basic_construction}, the remaining problem is how to construct explicit pooling matrix for \SCQGT{n,n^{\kappa},n^{\delta},n^{\lambda}}.
To construct such matrix with the desired pooling complexity, we propose an approach, called embedding method, to \emph{reduce} the original problem into constructing an \emph{unbalanced bipartite expander}, so that once the construction of the  bipartite expander is found, it leads to the construction of a $(n,n^{\kappa},\frac{\sqrt{1-\epsilon}}{2}n^\delta,n^\lambda)$-detecting matrix with height asymptotically being $n^{\lambda} \lp\frac{\lambda\log^2 n}{\epsilon}\rp^{3}4^{\sqrt{3\lambda \log n\log \lp\frac{\lambda\log^2 n}{\epsilon}\rp}} = o(n^{\lambda+\rho})$ for all $\rho>0$, satisfying the aforementioned desired requirement on the height of $\mb{P}$.

{
The high-level idea behind our embedding method goes as follows. If we can embed our construction for CQGT $\mb{Q}$ with width roughly equal to the sparsity level of the SCQGT problem in $\mb{P}$ such that whatever the underlying support set of hidden data $\bm{x}$ is, the corresponding column-submatrix always contain a row-submatrix equal to $\mb{Q}$, where column(row)-submatrix mean submatrix composed by certain columns(rows) of the original matrix. The constructed $\mb{P}$ would be a good pooling matrix for the SCQGT problem because whatever the support of the underlying data is, it always sees $\mb{Q}$, a good pooling matrix for CQGT. Consequently, we reduce the SCQGT problem to the CQGT problem non-adaptively. However, this exact embedding criteria is unlikely to reach. Since we need to make sure that all $\tbinom{n}{2n^{\lambda}}$ submatrices $\mb{P}_{2n^{\lambda}}$ satisfy this criterion. The key ingredient of our embedding method is that we can relax this exact embedding criterion if we have a robust pooling matrix for the CQGT problem. Suppose we carefully design a robust pooling matrix $\mb{H}$ such that it remains a pooling matrix for the CQGT problem even if part of the matrix is perturbed arbitrarily. In that case, we can largely relax our embedding criteria since we only need $\mb{P}_{2n^{\lambda}}$ looks roughly like $\mb{H}$ instead of exactly equal to $\mb{H}$. Once the difference between them does not exceed the robustness level of $\mb{H}$, $\mb{P}$ would remain a good pooling matrix for the SCQGT problem. It turns out that Hadamard matrices with size $2n^{\lambda}$ are a good candidate for $\mb{H}$ and we can perform our embedding by leveraging a combinatorial structure called \emph{unbalanced bipartite expanders}.}
%The high-level intuition behind this reduction is straightforward. We already knew that the Hadamard matrix is good at dealing with infinity norm noise. The only disadvantage of it is high pooling complexity. Therefore, if we can efficiently embed small Hadamard structures in our pooling matrix such that every sparse vector we are concerned would see a Hadamard-like structure in submatrix induced by its non-zero entries, then the criteria can be satisfied.} To describe how the reduction works, let us first introduce the notions of bipartite graphs and bipartite expanders.

\begin{definition}[Bipartite Graphs]
A bipartite graph $\mcal{G} = (\msf{L},\msf{R},\msf{E})$ consists of the set of \emph{left} vertices $\msf{L}$, the set of \emph{right} vertices $\msf{R}$, and the set of edges $\msf{E}$, where 
each edge in $\msf{E}$ is a pair $(i,j)$ with $i \in \msf{L}$ and $j\in\msf{R}$. Note that we follow the convention that left and right vertex sets are equivalently represented by their index sets, that is, the following equivalence: 
\[
\msf{L} \equiv [N] := \{1,...,N\},\quad \msf{R} \equiv [M] := \{1,...,M\},
\]
where $|\msf{L}| = N$ and $|\msf{R}| = M$. 

A bipartite graph is called \emph{left-$D$-regular} if each left vertex $i \in \msf{L}$ has exactly $D$ neighbors in the right part $\msf{R}$, that is, the cardinality of the \emph{neighbor} of $i$ (denoted by $\Gamma(i)$) is exactly $D$ for all $i \in \msf{L}$
\end{definition}

\begin{definition}[Bipartite Expanders]
A bipartite, left-$D$-regular graph $\mcal{G} = ([N],[M],\msf{E})$ is called a $(N,M,D,K,A)$-bipartite expander if $\forall\, S\subseteq [N]$ with $|S|\leq K$, 
\[
|\Gamma(S)|\geq A|S|,\ 
\text{where $\Gamma(S) := \bigcup_{i\in S} \Gamma(i)$, the union of all neighbors of nodes in $S$.}
\]
%Where $[N]$ correspond to $N$ left node, $[M]$ correspond to $M$ right node, $[D]$ correspond to $D$ edges for each left node, and $\Gamma(S)$ is the shorthand of $\Gamma(S,[D])$, the set of all neighbors of $S$.
%A left $D$ regular bipartite graph $\Gamma : [N]\times [D]\rightarrow [M]$ is called a $(K,A)$ bipartite expander if $\forall S\subset [N], |S|\leq K, |\Gamma(S)|\geq A|S|$. Where $[N]$ correspond to $N$ left node, $[M]$ correspond to $M$ right node, $[D]$ correspond to $D$ edges for each left node, and $\Gamma(S)$ is the shorthand of $\Gamma(S,[D])$, the set of all neighbors of $S$.
\end{definition}

\begin{figure}[htbp]
    \centering
    \includegraphics[width=0.6\linewidth]{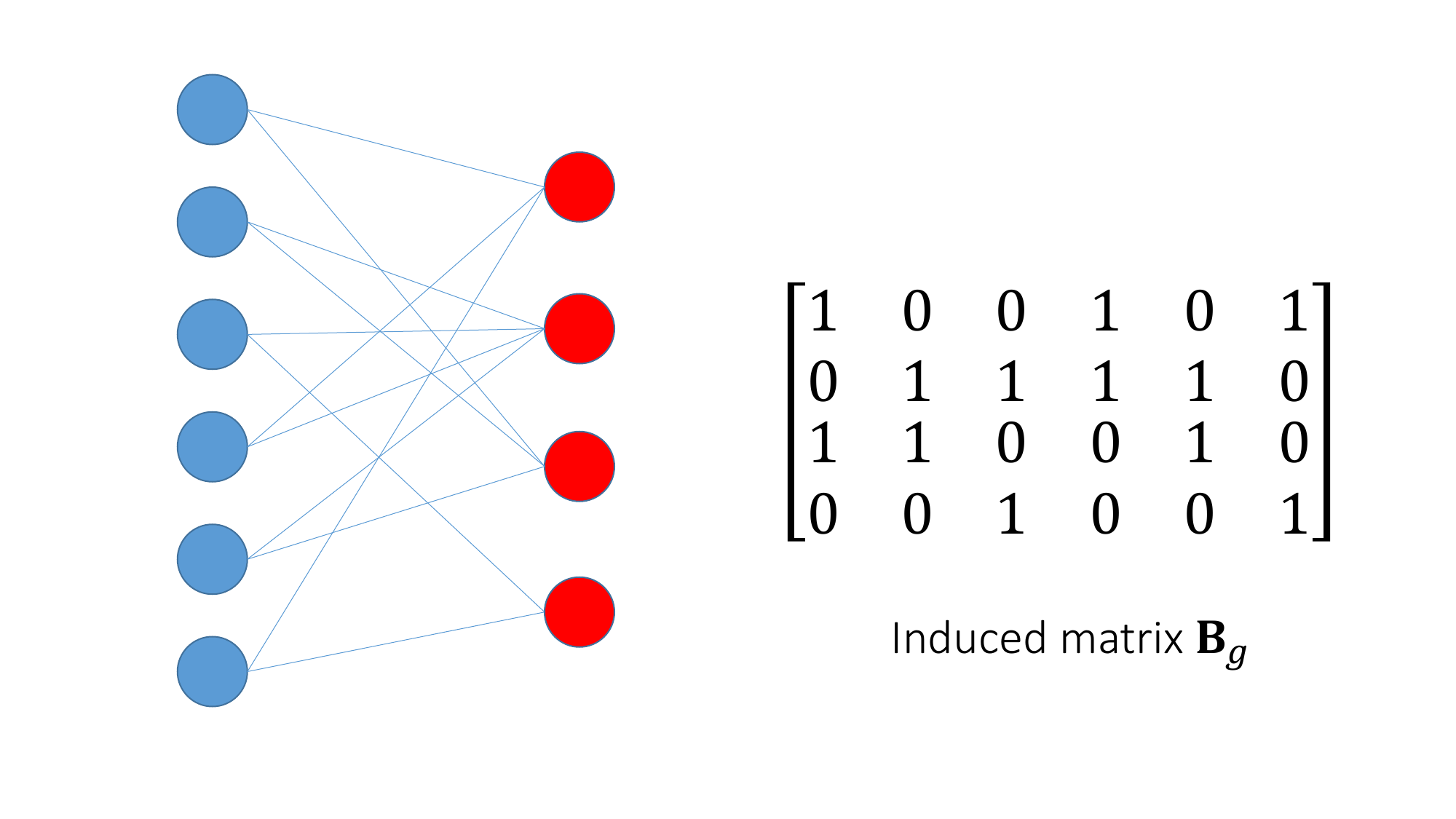}
    \caption{A $(6,4,2,2,1)$-bipartite expander and its induced matrix $\mb{B}_{\mcal{G}}$.}
    \label{fig:expander_label}
\end{figure}

Before we proceed, let us give an example of a $(6,4,2,2,1)$-bipartite expander in Figure~\ref{fig:expander_label}.

%{\color{blue}Here, we give an example of bipartite expander in Figure \ref{fig:expander_label}, where it is a (2,1) bipartite expander with $N=6$ left nodes, $M=3$ right nodes, and left degree $D=2$.}

In our construction, we heavily rely on the binary matrix (denoted by $\mb{B}_{\mcal{G}}$) induced by a bipartite graph $\mcal{G} = (\msf{L},\msf{R},\msf{E})$, with its $(i,j)$-th entry being 
\[
(\mb{B}_{\mcal{G}})_{i,j} := 
\begin{cases} 
1, &\text{if $(i,j)\in\msf{E}$;}\\
0, &\text{otherwise.}
\end{cases}
\]
Hence, by definition, for a $(N,M,D,K,A)$-bipartite expander $\mcal{G}$, its corresponding induced binary matrix $\mb{B}_{\mcal{G}}$ satisfies the following two properties:
\begin{itemize}
    \item Each of its column vector is $D$-sparse, that is, each of them has exactly $D$ $1$'s.
    \item For any $k$ of its column vectors $\bm{B}_{i_1},...,\bm{B}_{i_k}$, $1\leq k\leq K$, $\lVert\bigvee_{j = i_{1},...,i_{k}}\bm{B}_{j}\rVert_{0}\geq Ak$, where $\bigvee$ denotes the bit-wise ``or'' operation of binary column vectors.
\end{itemize}

%%%%%%
%%%%%%

\begin{figure}[htbp]
    \centering
    \includegraphics[width = \linewidth]{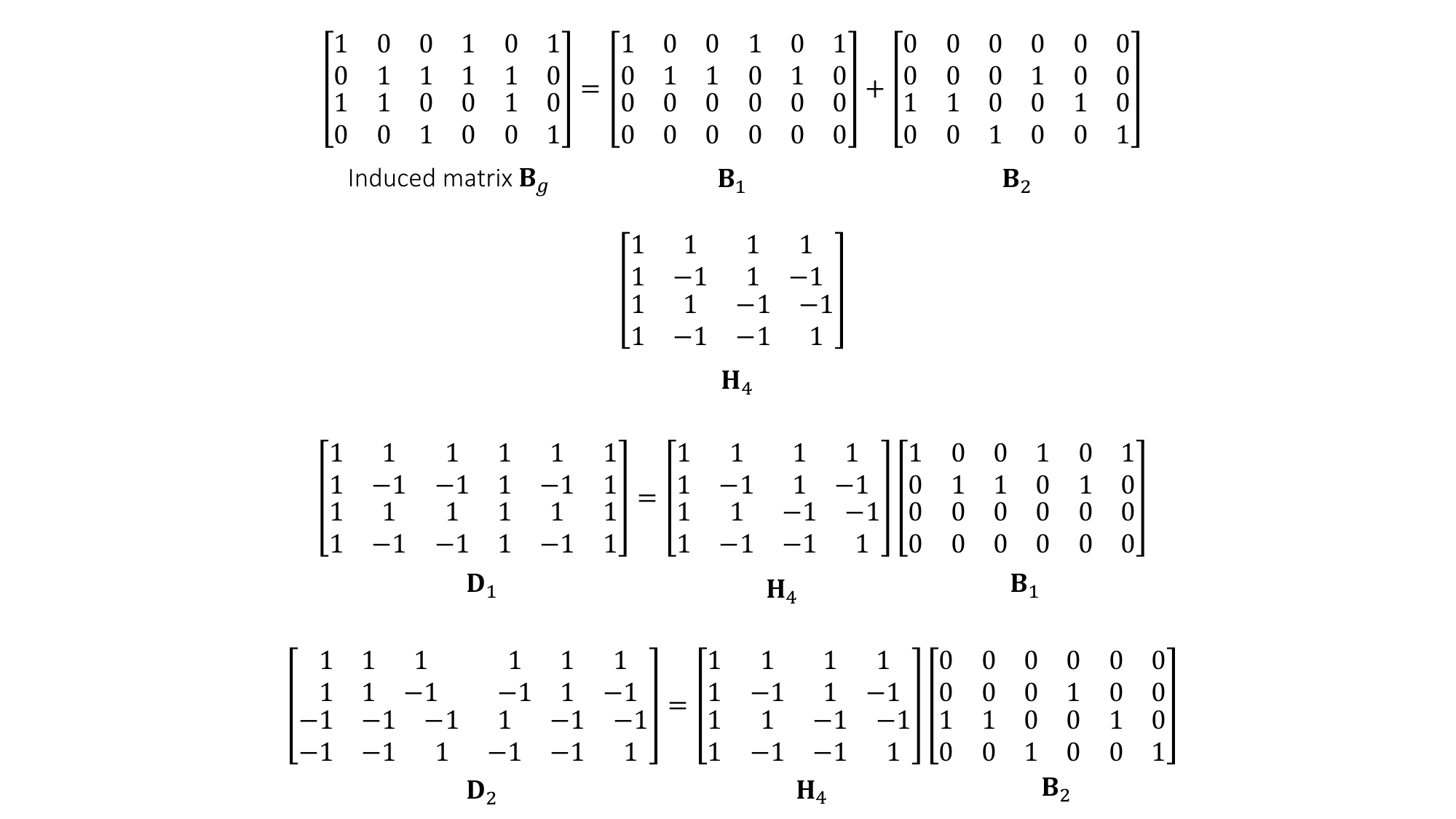}
    \caption{example of the reduction procedure}
    \label{fig:my_label}
\end{figure}

We are now ready to describe the proposed reduction. Consider a $(N,M,D,K,A)$-bipartite expander $\mcal{G}$, with its parameters $(N,M,D,K,A)$ to be specified later. The proposed pooling matrix $\mb{P}$ takes the following form: 
\begin{equation}\label{eq:design_of_Q}
\mb{P} = \begin{bmatrix}\mb{D}_{1}\\  \vdots \\ \mb{D}_{D}\end{bmatrix},
\end{equation}
where $\mb{D}_{1},...,\mb{D}_{D}$ are block matrices to be constructed from the expander $\mcal{G}$, as described in the following steps:
\begin{enumerate}
\item 
First, take the induced binary matrix $\mb{B}_{\mcal{G}}$ of the $(N,M,D,K,A)$-bipartite expander $\mcal{G}$.
\item 
Second, decompose $\mb{B}_{\mcal{G}}$ into $D$ binary matrices $\mb{B}_{1},...,\mb{B}_{D}$, such that each of their columns has exactly one non-zero element, and 
\[
\sum_{i=1}^{D}\mb{B}_{i} = \mb{B}_{\mcal{G}}.
\]
\item 
Third, for $i=1,...,D$, construct the $i$-th block matrix $\mb{D}_i$ as
\[
\mb{D}_{i} = \mb{H}_{M}\mb{B}_{i},
\]
where $\mb{H}_{M}$ is the Hadamard matrix of size $M$. 
\end{enumerate}
Let us illustrate the construction with an example depicted in Figure~\ref{fig:my_label}. In this example, we first decompose $\mb{B}_{\mcal{G}}$ into $\mb{B}_{1}+\mb{B}_{2}$, and then construct $\mb{D}_{1} = \mb{H}_{4}\mb{B}_{1}$ and $\mb{D}_{2} = \mb{H}_{4}\mb{B}_{2}$ respectively.

To this end, what remains is how to choose the parameters $(N,D,M,K,A)$ in the bipartite expander so that the constructed pooling matrix $\mb{P}$ in \eqref{eq:design_of_Q} meets our need. 
It turns out that with the explicit construction of bipartite expanders by Guruswami \textit{et al.} \cite{expander_construction}, we are able to choose 
\begin{equation}\label{eq:chosen_parameter}
\text{$(N,D,M,K,A) = \lp n, \frac{\lambda \log(n)^{2}}{\epsilon}2^{\sqrt{\frac{\lambda \log(n) \log\lp \frac{\lambda \log(n)^{2}}{\epsilon}\rp }{3}}},  D^{2}n^{\lambda}2^{\sqrt{3\lambda \log(n) \log\lp \frac{\lambda \log(n)^{2}}{\epsilon}\rp }}, n^{\lambda}, (1-\epsilon)D\rp$ }
\end{equation}
in the explicit construction, and together with the reduction mentioned above, the desired pooling matrix $\mb{P}$ can be obtained. The following theorem summarizes the guarantees of the constructed $\mb{P}$.   

\begin{theorem}\label{final_form}
With the above construction, the pooling matrix $\mb{P}$ in equation \eqref{eq:design_of_Q} is $(n,n^{\kappa},\frac{\sqrt{1-2\epsilon}}{2}n^\delta,n^\lambda)$-detecting and its pooling complexity is at most $n^{\lambda} \lp\frac{\lambda\log(n)^{2}}{\epsilon}\rp^{3}4^{\sqrt{3\lambda \log\lp n\rp\log \lp\frac{\lambda\log(n)^{2}}{\epsilon}\rp}}$.
\end{theorem}
\begin{IEEEproof}

{Recall that form of the constructed matrix $\mb{P}$ in \eqref{eq:design_of_Q}.
For each $\mb{D}_{i}$, 
\begin{align*}
    \forall \bm{x},\bm{x}' &\in \{0,1\}^{n\times 1}, \lVert\bm{x}\rVert_0 , \lVert\bm{x}'\rVert_0 < n^{\lambda}, \\
    &\lnorm\mb{D}_{i}\lp \bm{x}-\bm{x}'\rp\rnorm_{2}^{2} = \lnorm\mb{H}_{M}\mb{B}_{i}\lp \bm{x}-\bm{x}'\rp\rnorm_{2}^{2} =^{(a)} M\lnorm \mb{B}_{i}\lp \bm{x}-\bm{x}'\rp\rnorm_{2}^{2}\geq^{(b)} M\lvert S_{\bm{x}-\bm{x}'}^{i}\rvert
\end{align*}
where $S_{\bm{x}-\bm{x}'}^{i}$ is defined as follows. Let $\mathrm{supp}\lp \bm{x}-\bm{x}'\rp$ be the support set of ternary vector $\bm{x}-\bm{x}'$. We then define 
\begin{equation*}
    S_{\bm{x}-\bm{x}'}^{i} := \{j\in \mathrm{supp}\lp \bm{x}-\bm{x}'\rp \lvert  \forall j^{'}\neq j\in \mathrm{supp}\lp \bm{x}-\bm{x}'\rp, \mb{B}_{i}[j]\neq \mb{B}_{i}[j^{'}]\}    
\end{equation*}
$\mb{B}_{i}[j]$ denote the $j$-th column of matrix $\mb{B}_{i}$.  

$(a)$ follows from the property of Hadamard matrix $\lp\forall \bm{x}\in \mathbb{R}^{M}, \lVert\mb{H}_{M}\bm{x}\rVert_{2}^{2} = M\lVert \bm{x}\rVert_{2}^{2}\rp$. $(b)$ follows from the definition of $S_{\bm{x}-\bm{x}'}^{i}$, and the fact that all column vector of $\mb{B}_{i}$ are unit vectors, consequently there would be no two column vectors with indices in $S_{\bm{x}-\bm{x}'}^{i}$ contribute to the same element in $\mb{B}_{i}\lp \bm{x}-\bm{x}'\rp$. Combined with the fact that $\bm{x}-\bm{x}'$ is a ternary vector, each indices in $S_{\bm{x}-\bm{x}'}^{i}$ would correspond to an element in $\mb{B}_{i}\lp \bm{x}-\bm{x}'\rp$ with absolute value equal to $1$.    

Then 
\begin{align*}
    \lnorm\mb{P}\lp\bm{x}-\bm{x}'\rp\rnorm_{2}^{2} &= \sum_{i=1}^{D}\lnorm\mb{D}_{i}\lp \bm{x}-\bm{x}'\rp\rnorm_{2}^{2}\geq \sum_{i=1}^{D}M\lvert S_{\bm{x}-\bm{x}'}^{i}\rvert\\
    &\geq^{(a)} M\lp A\lnorm \bm{x}-\bm{x}'\rnorm_{0} - \lp D-A\rp \lnorm \bm{x}-\bm{x}'\rnorm_{0}\rp = M\lp 2A-D\rp \lnorm \bm{x}-\bm{x}'\rnorm_{0}
\end{align*}
$(a)$ follows from $\mb{B}_{\Gamma}$ is an adjacency matrix of $(N,M,D,K,A)$-bipartite expander graph, the fact that $\bm{x}-\bm{x}'$ is a ternary vector, and the fact that if $\mb{B}_{\Gamma}\lp \mathrm{supp} \lp\bm{x}-\bm{x}'\rp\rp>A\lvert \mathrm{supp}\lp\bm{x}-\bm{x}'\rp \rvert$, then there must be at least $\lp 2A-D\rp\lvert \mathrm{supp}\lp\bm{x}-\bm{x}'\rp \rvert$ right nodes connect to only one left node in $\mathrm{supp}\lp\bm{x}-\bm{x}'\rp$.  

Then
\begin{align*}
    \lnorm\mb{P}\lp\bm{x}-\bm{x}'\rp\rnorm_{\infty}&\geq^{(a)}\sqrt{\frac{\lnorm\mb{P}\lp\bm{x}-\bm{x}'\rp\rnorm_{2}^{2}}{\text{size of }\mb{P}}} \geq \sqrt{\frac{ M\lp 2A-D\rp \lnorm \bm{x}-\bm{x}'\rnorm_{0}}{DM}}\\
    &=\sqrt{\frac{2A-D}{D}\lnorm \bm{x}-\bm{x}'\rnorm_{0}}\geq^{(b)} \sqrt{\frac{2A-D}{D}}n^{\frac{\kappa}{2}}\geq^{(c)} \sqrt{\frac{2A-D}{D}}n^{\delta}
\end{align*}

$(a)$ follows from max always bigger than mean.$(b),(c)$ follows from the context of $(n,n^{\kappa},\sqrt{1-2\epsilon}n^\delta,n^\lambda)$-SCQGT problem. \\
For pooling complexity,
\begin{equation*}
    \text{size of matrix }\mb{P} = \sum_{i=1}^{D}\text{size of matrix }\mb{D}_{i} = MD    
\end{equation*}

The remaining problem is how to choose good parameter $(N,D,M,K,A)$ and how to construct such bipartite expander explicitly. It turns out we can obtain explicit construction of bipartite expander from \cite{expander_construction} Theorem 3.5 with parameter $N=n, D = \frac{\lambda \log(n)^{2}}{\epsilon}2^{\sqrt{\frac{\lambda \log(n) \log\lp \frac{\lambda \log(n)^{2}}{\epsilon}\rp }{3}}}, M = D^{2}n^{\lambda}2^{\sqrt{3\lambda \log(n) \log\lp \frac{\lambda \log(n)^{2}}{\epsilon}\rp }}, K=n^{\lambda}, A = (1-\epsilon)D$. Finally, plug in these specific parameter, the proof is complete.}
\end{IEEEproof}

%do we obtain $\mb{B}_{\Gamma}$ with suitable parameter($N,D,M,K,A$)? It turns out that by the explicit construction of bipartite expander in \cite{expander_construction} with suitable parameter, and the reduction mentioned above, we can obtain our desire matrix $\mb{Q}$.  
%The following theorem summarizes the guarantees of the constructed $\mb{Q}$, and its proof can be found in  Appendix~\ref{proof_final_form} of the extended version.  
%\begin{theorem}\label{final_form}
%The constructed matrix $\mb{Q}$ in equation \eqref{eq:design_of_Q} is $(n,n^{\kappa},\frac{\sqrt{1-2\epsilon}}{2}n^\delta,n^\lambda)$-detecting and its pooling complexity is at most $n^{\lambda} \lp\frac{\lambda\log(n)^{2}}{\epsilon}\rp^{3}4^{\sqrt{3\lambda \log\lp n\rp\log \lp\frac{\lambda\log(n)^{2}}{\epsilon}\rp}}$.
%\end{theorem}
%\begin{align*}
%    &\forall \bm{x},\bm{x}' \in \{0,1\}^{n\times 1}, \lVert\bm{x}\rVert_0 , \lVert\bm{x}'\rVert_0 < l_n, \\
%    &\textstyle\lVert \mb{Q}\bm{x} - \mb{Q}\bm{x}'\rVert_\infty > 
%    \frac{\lVert \bm{x}-\bm{x}'\rVert_{2}}{48\log(n)^{\frac{3}{2}}}
%\end{align*}

\subsection{Decoding algorithms and complexities}\label{subsec:decoding}

Let us turn to the decoding algorithms for measurement plan mentioned in Theorem \ref{basic_construction} in Section~\ref{Basic_tmp}, and decoding algorithms for measurement plan mentioned in Theorem \ref{final_form} in Section~\ref{SCQGT_decoding_tmp}.      .

\subsubsection{Decoding algorithm for Basic code}\label{Basic_tmp}
We propose a two step $O\lp n\rp $ decoding algorithm for the basic construction:
\begin{enumerate}
    \item Deconstruction step
    \item Rounding step
\end{enumerate}

For the deconstruction step, let $\bm{y}^{'} = \mb{Q}\bm{x^{'}}+\bm{n}^{'}, \bm{x^{'}}\in\{0,1\}^{n}$. Since $\mb{Q}$ is reduced from $\mb{Q}_{\bar{n}}^{3}$, $\bm{y}^{'} = \mb{Q}_{\bar{n}}^{3}\bm{x}+\bm{n}^{'}, \bm{x} = \lb \bm{x^{'}}, \bm{0}_{\bar{n} - n} \rb\in \{0,1\}^{\bar{n}}$, where $\bm{0}_{\bar{n} - n}$ is the zero vector with size $\bar{n} - n$. Let $\epsilon = \kappa-2\delta$. We first subtract $\bm{y}_{l}^{'}$(lower half of $\bm{y}^{'}$) from $\bm{y}_{u}^{'}$(upper half $\bm{y}^{'}$), then
\begin{equation*}
    \bm{y} = \bm{y}_{u}^{'}-\bm{y}_{l}^{'} = \lp \mb{Q}_{\bar{n}}^{1}-\mb{Q}_{\bar{n}}^{2}\rp \bm{x} + \bm{n}_{u}^{'}-\bm{n}_{l}^{'} = \mb{Q}_{\bar{n}}^{0}\bm{x}+\bm{n}
\end{equation*}
Note that Sylvester's type Hadamard matrix with size $2^d$ can be write as kronecker product of $\mb{H}_2$ itself $d$ times
\begin{equation*}
    \mb{H}_{2^d} = \mb{H}_{2}\otimes \mb{H}_{2} \otimes ... \otimes \mb{H}_2 = \mb{H}_{2}^{\otimes d}
\end{equation*}
hence

\begin{align*}
    \mb{Q}_{\bar{n}}^{0} &= \mb{H}_{\slceil n^{1-\frac{\epsilon}{2}}\srceil } \otimes \mb{M}_{\slceil n^{\frac{\epsilon}{2}} \srceil } = \mb{H}_2 \otimes \mb{H}_{\frac{\slceil n^{1-\frac{\epsilon}{2}}\srceil }{2}} \otimes \mb{M}_{\slceil n^{\frac{\epsilon}{2}} \srceil } \\
    &= \mb{H}_2 \otimes  \mb{Q}_{\frac{\bar{n}}{2}}^{0}= 
        \lp 
            \begin{array}{cc}
                \mb{Q}_{\frac{\bar{n}}{2}}^{0}  & \mb{Q}_{\frac{\bar{n}}{2}}^{0}\\
                \mb{Q}_{\frac{\bar{n}}{2}}^{0}  & -\mb{Q}_{\frac{\bar{n}}{2}}^{0}
            \end{array}
        \rp
\end{align*}

then we can see that
\begin{equation*}
    \bm{y}
    =\lp
        \begin{array}{c}
            \bm{y}_{u}\\
            \bm{y}_{l}
        \end{array}
    \rp
    =
    \lp
            \begin{array}{cc}
                \mb{Q}_{\frac{\bar{n}}{2}}^{0}  & \mb{Q}_{\frac{\bar{n}}{2}}^{0}\\
                \mb{Q}_{\frac{\bar{n}}{2}}^{0}  & -\mb{Q}_{\frac{\bar{n}}{2}}^{0}
            \end{array}
        \rp \bm{x} + \bm{n}
\end{equation*}
Next we do some row operation(deconstruction)% for any column vector $A$ with even size, we define $R(A,1) = [\frac{A_{u}+A_{l}}{2}, \frac{A_{u}-A_{l}}{2}]^{T}$, where $A_{u}, A_{l}$ is the upper half and lower half of $A$. And $$ 
\begin{equation*}
    \lp
        \begin{array}{c}
            \frac{\bm{y}_{u}+\bm{y}_{l}}{2}\\
            \frac{\bm{y}_{u}-\bm{y}_{l}}{2}
        \end{array}
    \rp
     = 
     \lp
        \begin{array}{cc}
            \mb{Q}_{\frac{\bar{n}}{2}}^{0}  & \mb{0}                \\
            \mb{0}                 & \mb{Q}_{\frac{\bar{n}}{2}}^{0}
        \end{array}
    \rp
    \bm{x} + 
    \lp
        \begin{array}{c}
            \frac{\bm{n}_{u}+\bm{n}_{l}}{2}\\
            \frac{\bm{n}_{u}-\bm{n}_{l}}{2}
        \end{array}
    \rp
\end{equation*}
After some calculation

\begin{align*}
    \lnorm 
    \lp
        \begin{array}{c}
            \frac{\bm{n}_{u}+\bm{n}_{l}}{2}\\
            \frac{\bm{n}_{u}-\bm{n}_{l}}{2}
        \end{array}
    \rp
    \rnorm ^2 = \lnorm \frac{\bm{n}_{u}+\bm{n}_{l}}{2}\rnorm ^{2} + \lnorm \frac{\bm{n}_{u}-\bm{n}_{l}}{2}\rnorm ^{2}
    = \frac{\lnorm \bm{n}_{u}\rnorm ^2+\lnorm \bm{n}_{l}\rnorm ^2}{2} = \frac{\lnorm \bm{n}\rnorm ^2}{2}
\end{align*}

We can see that the two-norm square of noise vector reduce by half after one time deconstruction. Hence, after we do $\log_{2}\lp {\slceil n^{1-\frac{\epsilon}{2}}\srceil }\rp$ times deconstruction
\begin{align}\label{totally deconstruction}
    R\lp \bm{y}, \log_{2}\lp {\slceil n^{1-\frac{\epsilon}{2}}\srceil }\rp\rp =  
        \lp
            \begin{array}{cccc}
                \mb{Q}_{\slceil n^{\frac{\epsilon}{2}} \srceil }^{0}  & \mb{0}                             & ...       & \mb{0}                         \\
                \mb{0}                           & \mb{Q}_{\slceil n^{\frac{\epsilon}{2}} \srceil }^{0}    & ...       & \mb{0}                         \\
                \vdots                      & \vdots                        & \ddots    & \vdots                    \\
                \mb{0}                           & \mb{0}                             & ...       & \mb{Q}_{\slceil n^{\frac{\epsilon}{2}} \srceil }^{0} 
            \end{array}
        \rp
        \lp
            \begin{array}{c}
                \bm{x_1}\\
                \bm{x_2}\\
                \vdots\\
                \bm{x}_{\slceil n^{1-\frac{\epsilon}{2}}\srceil }
            \end{array}
        \rp 
        + R\lp \bm{n},\log_{2}\lp {\slceil n^{1-\frac{\epsilon}{2}}\srceil }\rp\rp 
\end{align}
Here we define $R(\bm{y},t), R(\bm{n},t)$ as the corresponding vector after doing $t$ times deconstruction on column vector $\bm{y}, \bm{n}$.

Since $\lnorm \bm{n}\rnorm _{\infty}\leq n^{\delta}$, and $s = o\lp n\rp $
\begin{align}\label{noise_strength}
    \lnorm R\lp \bm{n},\log_{2}\lp {\slceil n^{1-\frac{\epsilon}{2}}\srceil }\rp\rp\rnorm ^{2} = \frac{\lnorm \bm{n}\rnorm ^{2}}{\slceil n^{1-\frac{\epsilon}{2}}\srceil } = \frac{o\lp n^{1+2\delta}\rp }{\slceil n^{1-\frac{\epsilon}{2}}\srceil } 
    = o\lp n^{2\delta+\frac{\epsilon}{2}}\rp  
\end{align}
Then we divide $R\lp \bm{y}, \log_{2}\lp {\slceil n^{1-\frac{\epsilon}{2}}\srceil }\rp\rp = [\bm{y}_1 , \bm{y}_2 ,..., \bm{y}_{\slceil n^{1-\frac{\epsilon}{2}}\srceil } ]$ , $R\lp \bm{n},\log_{2}\lp {\slceil n^{1-\frac{\epsilon}{2}}\srceil }\rp\rp = [\bm{n}_1 , \bm{n}_2 ,..., \bm{n}_{\slceil n^{1-\frac{\epsilon}{2}}\srceil } ]$ into equal length segment. From $\lp \ref{totally deconstruction}\rp $, we can see that
\begin{equation*}
    \forall i, \bm{y}_i = \mb{Q}_{\slceil n^{\frac{\epsilon}{2}} \srceil }^{0}\bm{x}_{i}+\bm{n}_{i} = \mb{M}_{\slceil n^{\frac{\epsilon}{2}} \srceil }\bm{x}_{i}+\bm{n}_{i}
\end{equation*}

For the rounding step, for each $\bm{y}_i$, we first do rounding, and then apply decoding algorithm for noiseless code mentioned in section 4 of \cite{cantor_mills_1966}.

Since we do rounding first, if $\lnorm \bm{n}_i\rnorm _{\infty} < \frac{1}{2}$, then after rounding, the noisy part in $\bm{y}_i$ will vanish, and so the decoding result $\hat {\bm{x}}_{i} = \bm{x}_i$, hence, for those $i$ such that $\hat{\bm{x}}_{i} \neq \bm{x}_i$, $\lnorm \bm{n}_i\rnorm ^{2} \geq \frac{1}{4}$. Combine with \eqref{noise_strength}, the number of segments that possible wrong is smaller than
\begin{equation*}
    \frac{\lnorm R\lp \bm{n}, \log_{2}\lp {\slceil n^{1-\frac{\epsilon}{2}}\srceil }\rp\rp\rnorm ^{2}}{\frac{1}{4}} = o\lp n^{2\delta+\frac{\epsilon}{2}}\rp 
\end{equation*}
Since there are $\slceil n^{\frac{\epsilon}{2}} \srceil $ bits in each segment, the total number of error bits must smaller than 
\begin{equation*}
    o\lp n^{2\delta+\frac{\epsilon}{2}}\rp \slceil n^{\frac{\epsilon}{2}} \srceil  = o\lp n^{2\delta+\epsilon}\rp  = o\lp n^{\kappa}\rp 
\end{equation*}
Finally, in each deconstruction step we takes $O\lp\frac{n}{\log_{2}\lp n\rp }\rp$ operations, and we do $\log_{2}\lp {\slceil n^{1-\frac{\epsilon}{2}}\srceil }\rp$ times of deconstruction, so the decoding complexity in deconstruction step is $O\lp n\rp $. For the rounding step, it is easy to check that the decoding complexity for each data segment $x_i$ is $O\lp \slceil n^{\frac{\epsilon}{2}} \srceil \rp $, and there are totally $\slceil n^{1-\frac{\epsilon}{2}}\srceil $ segments, hence the total decoding complexity for the rounding step is $O\lp\bar{n}\rp  = O\lp n\rp $, so the total decoding complexity for basic code is $O\lp n\rp $.

\subsubsection{Decoding algorithm for \SCQGT{n,n^{\kappa},n^{\delta},n^{\lambda}}}\label{SCQGT_decoding_tmp}
Let us turn to the decoding algorithms for measurement plan mentioned in Theorem \ref{final_form}.
\begin{theorem}\label{expander_decode_SCQGT}
There is a decoding algorithm for the non-adaptive measure plan for $(n,4n^{\kappa},\frac{\sqrt{1-2\epsilon}}{2}n^\delta,n^\lambda)$-SCQGT in Theorem \ref{final_form} with decoding complexity $O\lp n\frac{\lambda\lp\lambda-2\delta\rp \log(n)^{3}}{\epsilon}2^{\sqrt{\frac{\lambda \log(n) \log\lp \frac{\lambda \log(n)^{2}}{\epsilon}\rp }{3}}}\rp$ 
\end{theorem}
\begin{IEEEproof}
The decoding procedure goes as follows.
\begin{enumerate}
    \item Get the pooling result $\bm{y} = \begin{bmatrix}\bm{y}_{1}^{\intercal} &  ... & \bm{y}_{D}^{\intercal}\end{bmatrix}^{\intercal} = \mb{P}\bm{x} + \bm{n} = \begin{bmatrix}\lp\mb{D}_{1}\bm{x}+\bm{n}_{1}\rp^{\intercal} &  ... & \lp\mb{D}_{D}\bm{x}+\bm{n}\rp^{\intercal}\end{bmatrix}^{\intercal}$.
    \item Compute $\bar{\bm{y}_{i}} = \mb{H}_{M}^{-1}\bm{y}_{i} = \mb{H}_{M}^{-1}\lp\mb{D}_{i}\bm{x}+\bm{n}_{i}\rp = \mb{H}_{M}^{-1}\mb{H}_{M}\mb{B}_{i}\bm{x} + \frac{\mb{H}_{M}}{M}\bm{n}_{i} = \mb{B}_{i}\bm{x} + \bar{\bm{n}_{i}}$, and then $\bm{y}_{i}^{*} = \mb{B}_{i}\bm{x} + \bm{n}_{i}^{*}$ which denote the rounding of $\bar{\bm{y}_{i}}$. 
    \item Call the Sparse Matching Pursuit algorithm that is illustrated in Fig 2 of \cite{expander_algorithm} as a black box with input $\bm{y}^{*} = \begin{bmatrix}\lp\bm{y}_{1}^{*}\rp^{\intercal} &  ... & \lp\bm{y}_{D}^{*}\rp^{\intercal}\end{bmatrix}^{\intercal} = \begin{bmatrix}\mb{B}_{1}^{\intercal} &  ... & \mb{B}_{D}^{\intercal}\end{bmatrix}^{\intercal}\bm{x} + \begin{bmatrix}\lp\bm{n}_{1}^{*}\rp^{\intercal} &  ... & \lp\bm{n}_{D}^{*}\rp^{\intercal}\end{bmatrix}^{\intercal}$. Get the result $\bm{x}^{*}\in Z^{n}$ of Sparse Matching Pursuit algorithm with guarantee $\lnorm \bm{x}-\bm{x}^{*}\rnorm_{1} = O\lp \lnorm \begin{bmatrix}\lp\bm{n}_{1}^{*}\rp^{\intercal} &  ... & \lp\bm{n}_{D}^{*}\rp^{\intercal}\end{bmatrix}^{\intercal} \rnorm_{1}/D\rp$ from \cite{expander_algorithm} Theorem 1.
    \item Base on $\bm{x}^{*}$, we make some refinements. We set those elements of $\bm{x}^{*}$ that is smaller than $0$ to $0$, and those greater than $1$ to $1$. We use $\bm{z}^{*}$ to denote our final output.
\end{enumerate}

Note that $\bm{y}_{i}^{*}$ is the rounding of $\bar{\bm{y}_{i}}$, and elements of $\mb{B}_{i}, \bm{x}$ are all integer, together implies $\bm{n}_{i}^{*}$ are integer vector, and $\lnorm\bm{n}_{i}^{*}\rnorm_{2}^{2}\leq 4\lnorm\bar{\bm{n}_{i}}\rnorm_{2}^{2}$. Our noise model also implies $\lnorm\bm{n}\rnorm_{2}^{2} = \lp\sqrt{1-2\epsilon}n^\delta\rp^{2}\times\text{size of }\mb{P} = \lp 1-2\epsilon\rp n^{2\delta}MD$. Combine these two fact, we get
\begin{align*}
    \lnorm \begin{bmatrix}\lp\bm{n}_{1}^{*}\rp^{\intercal} &  ... & \lp\bm{n}_{D}^{*}\rp^{\intercal}\end{bmatrix}^{\intercal} \rnorm_{2}^{2} = \sum_{i=1}^{D}\lnorm \bm{n}_{i}^{*}\rnorm_{2}^{2} \leq 4\sum_{i=1}^{D}\lnorm \bar{\bm{n}_{i}}\rnorm_{2}^{2} = 4\sum_{i=1}^{D}\frac{\lnorm \mb{H}_{M}\bm{n}_{i}\rnorm_{2}^{2}}{M^{2}} = 4\sum_{i=1}^{D}\frac{\lnorm \bm{n}_{i}\rnorm_{2}^{2}}{M} = \frac{4\lnorm \bm{n}\rnorm_{2}^{2}}{M}\leq 4\lp 1-2\epsilon\rp n^{2\delta}D
\end{align*}
since $\begin{bmatrix}\lp\bm{n}_{1}^{*}\rp^{\intercal} &  ... & \lp\bm{n}_{D}^{*}\rp^{\intercal}\end{bmatrix}^{\intercal}$ is an integer vector, we get
\begin{equation*}
    \lnorm \bm{x}-\bm{x}^{*}\rnorm_{1} = O\lp \lnorm \begin{bmatrix}\lp\bm{n}_{1}^{*}\rp^{\intercal} &  ... & \lp\bm{n}_{D}^{*}\rp^{\intercal}\end{bmatrix}^{\intercal} \rnorm_{1}/D\rp = O\lp \lnorm \begin{bmatrix}\lp\bm{n}_{1}^{*}\rp^{\intercal} &  ... & \lp\bm{n}_{D}^{*}\rp^{\intercal}\end{bmatrix}^{\intercal} \rnorm_{2}^{2}/D\rp = O\lp 4\lp 1-2\epsilon\rp n^{2\delta}\rp
\end{equation*}

In step 4, since $\bm{x}$ is a binary vector, 
\begin{equation*}
    \lnorm \bm{x}-\bm{z}^{*}\rnorm_{0}\leq \lnorm \bm{x}-\bm{x}^{*}\rnorm_{1} = O\lp 4\lp 1-2\epsilon\rp n^{2\delta}\rp = O\lp 4\lp 1-2\epsilon\rp n^{2\delta}\rp = O\lp 4\lp 1-2\epsilon\rp n^{\kappa}\rp
\end{equation*}
Hence the output $\bm{z}^{*}$ satisfy the SCQGT guarantee.

For decoding complexity, note that the decoding complexity for step 2 is $O\lp MD\log(M)\rp = O\lp n\rp$, for step 3, by \cite{expander_algorithm} Theorem 1, is 
\begin{equation*}
    O\lp nD\log\lp D\lnorm \bm{x}\rnorm_{1}/\lnorm \begin{bmatrix}\lp\bm{n}_{1}^{*}\rp^{\intercal} &  ... & \lp\bm{n}_{D}^{*}\rp^{\intercal}\end{bmatrix}^{\intercal} \rnorm_{1}\rp\rp = O\lp n\frac{\lambda\lp\lambda-2\delta\rp \log(n)^{3}}{\epsilon}2^{\sqrt{\frac{\lambda \log(n) \log\lp \frac{\lambda \log(n)^{2}}{\epsilon}\rp }{3}}}\rp
\end{equation*}
for step 4 is $O(n)$, hence the overall decoding complexity is dominated by step 3.
\end{IEEEproof}

\subsection{Explicit construction of the optimal CQGT pooling matrix and decoding algorithm}\label{subsec:final_form}
We are now ready to complete the program outlined in Section~\ref{Basic_construction}, that is, to combine the non-adaptive measurement plan in Section~\ref{Basic_construction} and the non-adaptive SCQGT pooling matrix in Section~\ref{adaptive_embedding} to produce an non-adaptive scheme for the original CQGT problem. 
In particular, the following theorem summarizes the detecting capability of the constructed pooling matrix $\mb{R}$ in \eqref{decompose}.

\begin{theorem}
In \eqref{decompose}, if the submatrix $\mb{Q}$ is a \CQGT{n,n^{\lambda},n^{\delta}} detecting matrix based on the construction in Section~\ref{Basic_construction} and $\mb{P}$ is a \SCQGT{n,n^{\kappa},n^{\delta},n^{\lambda}} detecting matrix based on the construction in Section~\ref{adaptive_embedding}, then the matrix $\mb{R}$ is a \CQGT{n,n^{\kappa},n^{\delta}} detecting matrix. Moreover, there is an efficient decoding algorithm with decoding complexity $$\frac{\lambda\lp\lambda-2\delta\rp}{\epsilon}O\lp n\log(n)^{3}2^{\sqrt{\frac{\lambda \log(n) \log\lp \frac{\lambda \log(n)^{2}}{\epsilon}\rp }{3}}}\rp$$.
\end{theorem}
\begin{IEEEproof}
%\CQGT{n,4n^{\kappa},\frac{\sqrt{1-2\epsilon}}{2}n^{\delta}} with pooling complexity $\frac{1}{1-2\delta}\frac{n}{\log_{2}n}$ to within a constant factor. The overall procedure goes as follows. 

\underline{Step A}: 
First, employ the measurement plan of the basic construction in Section~\ref{Basic_construction} for \CQGT{n,n^{\lambda},n^{\delta}} and use the companion decoding algorithm (described in Section~\ref{subsec:decoding}). 
%Appendix~\ref{decode algorithm}. 
The decoded result is then represented as follows:
\begin{equation}
\bm{\hat x} = \bm{x}-\bm{p}, 
\end{equation}
where $\bm{x}$ is the true data vector and $\bm{p}$ is $n^{\lambda}$-sparse $\{0,\pm1\}$-vectors. In this step, we make $s_{\text{A}} = \frac{48}{\lambda-2\delta}\frac{n}{\log_{2}n}$ counting measurements.

\underline{Step B}: 
In this step, $\bm{p}$, the remaining mistakes made in Step A, will be detected. 
Compute $\bm{\hat y} = \mb{P}\bm{\hat x}$, combined with $\bm{y} = \mb{P}\bm{x}+\bm{n}$ (part of the answer vector, where $\bm{n}$ denote the noise vector), we get $\bm{z}:= \bm{y}-\bm{\hat y} = \mb{P}\bm{p}+\bm{n}$. Then called the SCQGT decoding algorithm mentioned in Theorem \ref{expander_decode_SCQGT} to solve $\bm{p}$, and we denote it's output vector by $\bm{z}^{*}$. By Theorem \ref{expander_decode_SCQGT}, $\mb{P}$ is a detecting matrix for $(n,4n^{\kappa},\frac{\sqrt{1-2\epsilon}}{2}n^\delta,n^\lambda)$-SCQGT, and thus our final output $\bm{\hat x}+\bm{z}^{*}$ would differ from true data vector $\bm{x}$ at most $4n^{\kappa}$ bits. In this step, {we make $n^{\lambda} \lp\frac{\lambda\log(n)^{2}}{\epsilon}\rp^{3}4^{\sqrt{3\lambda \log\lp n\rp\log \lp\frac{\lambda\log(n)^{2}}{\epsilon}\rp}}$ counting measurements.}

Note that in Step B, $\bm{p}$ is ternary vector, not binary as supposed in Theorem \ref{expander_decode_SCQGT}, but it turns out that with little adjustment, the whole algorithm still work for ternary vector, and consequently the guarantee  in Step B follows. 

To sum up, the total number of mistakes made at the end is at most $4n^{\kappa}$ with the number of counting measurements no more than 
{\[\textstyle
s_{\text{total}} = s_{\text{A}}+s_{\text{B}} = \frac{48}{\lambda-2\delta}\frac{n}{\log_{2}n} + n^{\lambda} \lp\frac{\lambda\log(n)^{2}}{\epsilon}\rp^{3}4^{\sqrt{3\lambda \log\lp n\rp\log \lp\frac{\lambda\log(n)^{2}}{\epsilon}\rp}}.
\]}
Asymptotically, when $n$ is large enough, $s_{\text{total}} \approx \frac{48}{\lambda-2\delta}\frac{n}{\log_{2}n}$. Taking $\lambda \rightarrow 1$, $s_{\text{total}}$ tends to $\frac{48}{1-2\delta}\frac{n}{\log_{2}n}$. As a result, the total pooling complexity is $\frac{48}{1-2\delta}\frac{n}{\log_{2}n}$. 
Moreover, the total decoding complexity is dominated by Step B, the decoding algorithm in Theorem \ref{expander_decode_SCQGT}. Hence, the decoding complexity of the overall decoding algorithm is shown to be that claimed in the Theorem.
\end{IEEEproof}

\section{Proofs of the Fundamental Limits}\label{sec:fundamental_proof}
\subsection{Proof of Achievability (Lemma~\ref{non_constructive_proof} and \ref{sparse_achievability})}\label{subsec:pf_direct}

A probabilistic argument is used to show the existence of good pooling matrices. In particular, we are going to \emph{upper bound} the probability that a randomly generated matrix with height $s$ is \emph{not} an $(n,k_{n},d_{n})$-detecting matrix. If this probability is strictly bounded below $1$, then the existence of $(n,k_{n},d_{n})$-detecting matrices is established. In words, we are going to show that $s \le \frac{8}{1-2\delta}\frac{n}{\log n}$ is a sufficient condition for the upper bound mentioned above being strictly less than $1$. 

%by apply union bound on the probability that we cannot detect each possible difference vector $\bm{x}-\bm{\hat{x}}$, $\bm{x},\bm{\hat{x}}\in \{0,1\}^n,   \lnorm\bm{x}-\bm{\hat{x}}\rnorm _{0}\geq k_n$. If this upper bound smaller $1$ for some $s$, then there must be a $\lp n,k_{n},d_{n}\rp $-detecting matrix with height smaller than $s$.

%First, 

Let us now describe the random pooling matrix ensemble employed in this probabilistic argument. 
To simplify the analysis, we focus on pooling matrices with $\{\pm1\}$-entries. Note that any pooling vector with $\{\pm1\}$-entries can be generated by taking the difference of two pooling vectors with $\{0,1\}$-entries. Hence, at the end of our analysis, to conform with the original CQGT problem formulation, we need to double the pooling complexity upper bound. 
%We use random pooling. 
The random pooling matrix ensemble is generated as follows: each element of the matrix is drawn from $\{\pm1\}$ uniformly at random, i.i.d. across all entries. With a slight abuse of notation, let $\mb{R}$ denote this random matrix, that is,
\[
(\mb{R})_{i,j} \diid \Unif(\{\pm1\}),\ \forall\,(i,j) \in \{1,...,s\}\times\{1,...,n\},
\]
and let $\bm{R}_i$ denote the $i$-th row of $\mb{R}$. 
%(To simplify analysis, we use $\{-1,1\}$ rather than $\{0,1\}$, but one can always mimic a $\{-1,1\}$ measurement by the subtraction of two binary measurement). 

Consider the event $\mcal{E}$ that $\mb{R}$ is not an $(n,k_n,d_n)$-detecting matrix. By definition (Definition~\ref{def:CQGT}), 
\begin{equation}\label{eq:outage_event_1}
\mcal{E} = \lbp\!
\begin{array}{ll}
\exists\, \bm{x},\bm{x}' \in \{0,1\}^{n\times 1}\ \text{with}
&\lVert\bm{x} - \bm{x}'\rVert_0 > k_n\ \text{and}\\[1ex]
&\lVert \mb{R}\bm{x} - \mb{R}\bm{x}'\rVert_\infty \le 2d_n
\end{array}\!
\rbp.
\end{equation}
For notational convenience, let us introduce 
\begin{equation}\label{eq:difference_set}
D_{a}^{b} = \lbp \bm{x}-\bm{y} \,\mv\, \bm{x},\bm{y}\in\{0,1\}^{n\times 1}, a< \lVert \bm{x}-\bm{y}\rVert _{0}\leq b\rbp
\end{equation}
to denote the set of difference vectors of $\ell_1$-norm ranging from $a$ to $b$. With the notations above, the event $\mcal{E}$ can be succinctly written as 
\[\textstyle
\mcal{E} = \bigcup_{\bm{d}\in D_{k_n}^{n}} \big\{ \lVert\mb{R}\bm{d}\rVert_{\infty} \le 2d_n\big\}.
\]
Hence, by the Union Bound, 
\begin{align}\label{eq:ach_1}
    \Pr\{\mcal{E}\}
    \leq \sum_{\bm{d}\in D_{k_n}^{n}} \Pr\lbp \lVert\mb{R}\bm{d}\rVert_{\infty} \le 2d_n\rbp
    = \sum_{\bm{d}\in D_{k_n}^{n}}\prod_{i=1}^{s} \Pr\lbp \bm{R}_i\bm{d} \le 2d_n\rbp
\end{align}
Noting that the event $\{\bm{R}_i\bm{d} \le 2d_n\}$ is equivalent to the event that out of $\lVert\bm{d}\rVert_0$ i.i.d. $\Unif(\{\pm1\})$ random variables, the number of $+1$ and the number of $-1$ differ by at most $2d_n$, we have
\begin{align}\label{eq:ach_2}
\Pr\lbp \bm{R}_i\bm{d} \le 2d_n\rbp
= \sum_{\ell: \lVert\bm{d}\rVert_0-2d_n \le 2\ell \le \lVert\bm{d}\rVert_0+2d_n} \binom{\lVert\bm{d}\rVert_0}{\ell}2^{-\lVert\bm{d}\rVert_0}
\le 2d_n \binom{\lVert\bm{d}\rVert_0}{\lfl\frac{1}{2}\lVert\bm{d}\rVert_0\rfl}2^{-\lVert\bm{d}\rVert_0}
%&\overset{\aaaa}{\le} 2d_n\lp\pi\lce \frac{\lVert\bm{d}\rVert_0}{2}\rce\rp^{-1/2}
\overset{\aaaa}{\le} 2d_n\lp\pi\frac{\lVert\bm{d}\rVert_0}{2}\rp^{-1/2}
\end{align}
$\aaaa$ is due to the fact that $\binom{j}{\lfl j/2\rfl} \le 2^j\lp\pi j/2\rp^{-1/2}$ for all $j\in\mbb{N}$. 
Combining \eqref{eq:ach_1} and \eqref{eq:ach_2}, we get
\begin{align}
\Pr\{\mcal{E}\} 
\le \sum_{\bm{d}\in D_{k_n}^{n}} \lp 2d_n\sqrt{2/\pi}\rp^s\lVert\bm{d}\rVert_0^{-s/2}
= \sum_{\ell = \lfl k_n\rfl+1 }^{n} \lvert D_{\ell-1}^{\ell}\rvert  \lp 2d_n\sqrt{2/\pi}\rp^s\ell^{-s/2}
\label{eq:prob_upper_bd_1}
\end{align}

To proceed, the range of the above summation is divided into three regimes and bounded separately: the first regime is $k_{n}\leq \ell\leq k_{n}n^{2\epsilon}$, the second regime is $k_{n}n^{2\epsilon}\leq \ell\leq n^{1-2\epsilon}$, and the third regime is $n^{1-2\epsilon}\leq \ell\leq n$, where $\epsilon$ is a positive constant that is smaller than $(1-\log_n k_n)/4 = (1-\kappa)/4$. Then,
\begin{align}
%    \Pr\{\mcal{E}\}
%    &= \sum_{\ell = \lfl k_n\rfl+1 }^{n} \lvert D_{\ell-1}^{\ell}\rvert  \lp 2d_n\sqrt{2/\pi}\rp^s\ell^{-s/2}\\
\eqref{eq:prob_upper_bd_1}    
&\overset{\aaaa}{\leq} \lba D_{\lfl k_n\rfl}^{k_{n}n^{2\epsilon}}\rba  \lp \frac{2d_n}{\sqrt{k_n}}\sqrt{2/\pi}\rp^s \notag\\
    &\quad+ \lba D_{k_{n}n^{2\epsilon}}^{n^{1-2\epsilon}}\rba  \lp \frac{2d_n}{\sqrt{k_{n}n^{2\epsilon}}}\sqrt{2/\pi}\rp^s \notag\\
    &\quad+ \lba D_{n^{1-2\epsilon}}^{n}\rba  \lp \frac{2d_n}{\sqrt{n^{1-2\epsilon}}}\sqrt{2/\pi}\rp^s \notag\\
    &\overset{\bbbb}{\leq} \lp 2(n+1)\rp^{k_{n}n^{2\epsilon}}\lp \frac{2d_n}{\sqrt{k_{n}}}\sqrt{2/\pi}\rp^s \notag\\
    &\quad+ \lp 2(n+1)\rp^{n^{1-2\epsilon}} \lp \frac{2d_n}{\sqrt{k_{n}n^{2\epsilon}}}\sqrt{2/\pi}\rp^s \notag\\
    &\quad+ 3^{n} \lp \frac{2d_n}{\sqrt{n^{1-2\epsilon}}}\sqrt{2/\pi}\rp^s \notag\\
    &\overset{\cccc}{=} \lp 2(n+1)\rp^{n^{\kappa+2\epsilon}}\lp n^{\delta-\frac{\kappa}{2}}\sqrt{8/\pi}\rp^s \label{eq:prob_upper_bd_2a}\\
    &\quad+ \lp 2(n+1)\rp^{n^{1-2\epsilon}} \lp n^{\delta-\frac{\kappa}{2}-\epsilon}\sqrt{8/\pi}\rp^s \label{eq:prob_upper_bd_2b}\\
    &\quad+ 3^{n} \lp n^{\delta-\frac{1}{2}+\epsilon}\sqrt{8/\pi}\rp^s. \label{eq:prob_upper_bd_2c}
\end{align}
$\aaaa$ follows from dividing the whole summation into the three regimes mentioned above and applying the trivial lower bound of $\ell$ in each regime. $\bbbb$ follows from applying two different upper bounds on the sizes of difference sets:
\begin{align}
\label{eq:difference_set_bd1}
\lvert D_a^b\rvert 
&\textstyle
= \sum_{j=a+1}^b \binom{n}{j}2^j \le \sum_{j=0}^b \binom{n}{j} 2^b \le (n+1)^b2^b,\\
\notag%\label{eq:difference_set_bd2}
\lvert D_a^b\rvert 
&\textstyle
= \sum_{j=a+1}^b \binom{n}{j}2^j \le \sum_{j=0}^n \binom{n}{j} 2^j = 3^n.
\end{align}
$\cccc$ follows from plugging in $k_n = n^\kappa, d_n = n^\delta$. %, 0<2\delta\le\kappa < 1$. 

Finally, in order to ensure all the three terms \eqref{eq:prob_upper_bd_2a} -- \eqref{eq:prob_upper_bd_2c} vanish as $n\ra\infty$, since it is the most stringent to drive \eqref{eq:prob_upper_bd_2c} to zero, it suffices to choose 
\[
s = \frac{\log\lp 3\rp n}{\lp1/2-\delta-\epsilon\rp\log\lp n\rp+1}.
\] 
%Now, let us plug $s = \frac{2}{\frac{1}{2}-\delta-\epsilon}\frac{n}{\log(n)}+c$ for some constant $c$ into \eqref{eq:prob_upper_bd_2} and observe that 
%\begin{align*}
%
%\end{align*}
%\begin{align*}
%    1 &> \lp 2n\rp^{n^{\kappa+2\epsilon}}\lp n^{\delta-\frac{\kappa}{2}}\sqrt{2/\pi}\rp^s\\
%    &\quad + \lp 2n\rp^{n^{1-2\epsilon}} \lp n^{\delta-\frac{\kappa}{2}-\epsilon}\sqrt{2/\pi}\rp^s
%    + 3^{n} \lp n^{\delta-\frac{1}{2}+\epsilon}\sqrt{2/\pi}\rp^s
%\end{align*}
%Hence, there exists a $\{\pm1\}$-pooling matrix with size $s = \frac{\log 3}{\frac{1}{2}-\delta-\epsilon}\frac{n}{\log n}$, for all $\epsilon \in (0,\frac{1-\kappa}{4})$. 
Picking sufficiently small $\epsilon\in (0,\frac{1-\kappa}{4})$, we immediately see that it is also sufficient to choose 
\[
s = \frac{4}{1-2\delta}\frac{n}{\log n}
\]
to ensure \eqref{eq:prob_upper_bd_2a} -- \eqref{eq:prob_upper_bd_2c} all vanish as $n\ra\infty$. As a result, there exists a $\{\pm1\}$-pooling matrix with size $s = \frac{4}{1-2\delta}\frac{n}{\log n}$. 
%Let $\epsilon\rightarrow 0$, we get 
%$s = \frac{4}{1-2\delta}\frac{n}{\log\lp n\rp }$. 
Finally, note that a $\{0,1\}$-pooling matrix can be generated by simple row operations from $\{\pm1\}$-pooling matrix, with the increase of the height by at most a factor of $2$. Hence, there exists a binary pooling matrix with size $s = \frac{8}{1-2\delta}\frac{n}{\log n}$, and this completes the proof of Lemma~\ref{non_constructive_proof}.

As for the proof of achievability for the sparse case (Lemma~\ref{sparse_achievability}), we slightly modify the definition of event $\mcal{E}$ in \eqref{eq:outage_event_1} as follows:
%that $\mb{Q}$ is not an $(n,k_n,d_n,l_n)$-detecting matrix. Define \[
\[
\mcal{E} = \lbp\!
\begin{array}{ll}
\exists\, \bm{x},\bm{x}' \in \{0,1\}^{n\times 1}\ \text{with}
&\lVert\bm{x}\rVert_0 , \lVert\bm{x}'\rVert_0 < l_n\\
&\lVert\bm{x} - \bm{x}'\rVert_0 > k_n\ \text{and}\\[1ex]
&\lVert \mb{P}\bm{x} - \mb{P}\bm{x}'\rVert_\infty \le 2d_n
\end{array}\!
\rbp.
\]
In words, it is the event that $\mb{P}$ is not an $(n,k_n,d_n,l_n)$-detecting matrix. 
Then, following the same proof program, an upper bound on the probability of this event, similar to \eqref{eq:prob_upper_bd_1}, can be found as follows:
%\
%
%by the same argument as \eqref{eq:ach_1}, we get 
%\begin{align*}
%    \Pr\{\mcal{E}\}
%    &= \sum_{\bm{d}\in D_{k_n}^{l_n}}\prod_{i=1}^{s} \Pr\lbp \bm{R}_i\bm{d} \le d_n\rbp
%\end{align*}
%
%combine \eqref{eq:ach_2}, we get 
\begin{equation}\label{sparse_ach1}
    \Pr\{\mcal{E}\} 
    \le \sum_{\ell = \lfl k_n\rfl+1 }^{2l_n} \lvert \tilde{D}_{\ell-1}^{\ell}\rvert  \lp 2d_n\sqrt{2/\pi}\rp^s\ell^{-s/2}.
\end{equation}
Note that now in the definition of the difference set $\tilde{D}_a^b$, there is an additional condition $\lVert\bm{x}\rVert, \lVert\bm{y}\rVert \le l_n$, compared to that of $D_a^b$ in \eqref{eq:difference_set}.  
%similar as \eqref{eq:prob_upper_bd_1}.
Next, following the steps to upper bound \eqref{eq:prob_upper_bd_1} by \eqref{eq:prob_upper_bd_2a} -- \eqref{eq:prob_upper_bd_2c}, we derive an upper bound on \eqref{sparse_ach1} by diving the range of the above summation into three regimes and bounding them separately: the first regime is $k_{n}\leq \ell\leq k_{n}n^{2\epsilon}$, the second regime is $k_{n}n^{2\epsilon}\leq \ell\leq l_{n} n^{-2\epsilon}$, and the third regime is $l_{n} n^{-2\epsilon}\leq \ell\leq 2l_{n}$, where $\epsilon$ is a positive constant that is smaller than $(\log_n l_n-\log_n k_n)/4 = (\lambda-\kappa)/4$. Then, 
\begin{align}\label{sparse_ach2}
\eqref{sparse_ach1}
&\leq \lba D_{\lfl k_n\rfl}^{k_{n}n^{2\epsilon}}\rba  \lp \frac{2d_n}{\sqrt{k_n}}\sqrt{2/\pi}\rp^s \notag\\
    &\quad+ \lba D_{k_{n}n^{2\epsilon}}^{l_{n} n^{-2\epsilon}}\rba  \lp \frac{2d_n}{\sqrt{k_{n}n^{2\epsilon}}}\sqrt{2/\pi}\rp^s \notag\\
    &\quad+ \lba D_{l_{n} n^{-2\epsilon}}^{2l_{n}}\rba  \lp \frac{2d_n}{\sqrt{l_{n} n^{-2\epsilon}}}\sqrt{2/\pi}\rp^s \notag\\
    &\overset{\dddd}{\leq} \lp 2(n+1)\rp^{k_{n}n^{2\epsilon}}\lp \frac{2d_n}{\sqrt{k_{n}}}\sqrt{2/\pi}\rp^s \notag\\
    &\quad+ \lp 2(n+1)\rp^{l_{n} n^{-2\epsilon}} \lp \frac{2d_n}{\sqrt{k_{n}n^{2\epsilon}}}\sqrt{2/\pi}\rp^s \notag\\
    &\quad+ 2^{2l_{n}(\log_2(\frac{\e n}{2l_n})+1)} \lp \frac{2d_n}{\sqrt{l_{n} n^{-2\epsilon}}}\sqrt{2/\pi}\rp^s \notag\\
    &\overset{\eeee}{=} \lp 2(n+1)\rp^{n^{\kappa+2\epsilon}}\lp n^{\delta-\frac{\kappa}{2}}\sqrt{8/\pi}\rp^s \\
    &\quad+ \lp 2(n+1)\rp^{n^{\lambda-2\epsilon}} \lp n^{\delta-\frac{\kappa}{2}-\epsilon}\sqrt{8/\pi}\rp^s \label{sparse_ach3}\\
    &\quad+ 2^{2n^{\lambda}((1-\lambda)\log_2(n)+\log_2(\e))} \lp n^{\delta-\frac{\lambda}{2}+\epsilon}\sqrt{8/\pi}\rp^s .\label{sparse_ach4}
\end{align}
$\dddd$ follows from \eqref{eq:difference_set_bd1}, $\lvert \tilde{D}_a^b\rvert 
\le \sum_{j=0}^b \binom{n}{j}2^b$, and
\[\textstyle
\sum_{j=0}^b \binom{n}{j} \le \sum_{j=0}^b \frac{n^j}{j!} = \sum_{j=0}^b \frac{b^j}{j!}\lp\frac{n}{b}\rp^j \le \e^b\lp\frac{n}{b}\rp^b.
\]
%{\color{red}
%\begin{align*}
%    \frac{\tilde{D}_{a}^b}{2^b} &\le \sum_{j=0}^b \binom{n}{j} \le 2\binom{n}{b} = 2\frac{n!}{\lp n-b\rp ! b!} \\
%    &\le 4\frac{\sqrt{2\pi n}\lp\frac{n}{\e}\rp^n}{\sqrt{2\pi \lp n-b\rp}\lp\frac{\lp n-b\rp}{\e}\rp^{\lp n-b\rp}\sqrt{2\pi \lp b\rp}\lp\frac{\lp b\rp}{\e}\rp^{ b}}\\
%    &=4\sqrt{\frac{n}{2\pi\lp n-b\rp b}}\frac{n^n}{\lp n-b\rp^{n-b}b^b}\le \frac{n^n}{2\lp n-b\rp^{n-b}b^b}\\
%    &= \frac{1}{2}\lp\frac{n}{n-b}\rp^{n-b}\lp\frac{n}{b}\rp^b \le \lp\frac{\e n}{b}\rp^b
%\end{align*}
%
%This follows from $\lim_{n\rightarrow \infty}b = \infty$ and $b<<n$ and Stirling's approximation.
%}
%The only difference here is that we have different difference set $D$ in sparse case, thus we need to use different upper bound on the sizes of these difference sets at $(b)$. 
$\eeee$ follows from plugging in $l_n = n^\lambda, k_n = n^\kappa, d_n = n^\delta, 0<2\delta\le\kappa < \lambda < 1$. 

Then, following the same discussion in the non-sparse case, in order to make \eqref{sparse_ach2} -- \eqref{sparse_ach4} all vanish as $n\rightarrow \infty$, it suffices to choose a sufficiently small $\epsilon$ and 
\[
s = \begin{cases}
        \frac{2(1-\lambda)}{\lambda-2\delta}n^\lambda,    & 2\delta<\kappa\\
        \frac{2(1-\lambda)}{\lambda-2\delta}n^{\lambda}\log_{2} n, & 2\delta=\kappa
\end{cases}
\]
%Finally, note that a $\{0,1\}$-pooling matrix can be generated by simple row operations from $\{\pm1\}$-pooling matrix, with the increase of the height by at most a factor of $2$. Hence, there exists a binary pooling matrix with size 
%
As a result, there exists a $\{0,1\}$-pooling matrix with size
\[
s = \begin{cases}
        \frac{4(1-\lambda)}{\lambda-2\delta}n^\lambda,    & 2\delta<\kappa\\
        \frac{4(1-\lambda)}{\lambda-2\delta}n^{\lambda}\log_{2} n, & 2\delta=\kappa
\end{cases}
\]

\subsection{Proof of Converse (Lemma~\ref{lower_bound} and \ref{sparse_converse})}\label{subsec:pf_converse}
The proof of converse is based on packing. It will be shown that if a pooling matrix $\mb{R}$ is $(n,k_n,d_n)$-detecting, the number of measurements $s$ (the height of $\mb{R}$) must be greater than or equal to a certain threshold. 
The argument goes as follows. 
Note that the detection criterion \eqref{eq:detect_criterion} implies that, for any $k_n$-packing $C_G$ with respect to $\ell_1$-norm of a subset $G\subseteq \{0,1\}^n$, its image set after multiplying with $\mb{R}$, $\mb{R}[C_G] := \{\mb{R}\bm{x}\,\vert\,\bm{x}\in C_G\}$, must be a $2d_n$-packing with respect to $\ell_\infty$-norm of the image set $\mb{R}[G] := \{\mb{R}\bm{x}\,\vert\,\bm{x}\in G\}$. By properly choosing $G$, one can derive a good upper bound on the packing number of $\mb{R}[G]$ which is related to $s$, the height of $\mb{R}$. Meanwhile, a lower bound of the packing number of $G$ is also a lower bound of the packing number of $\mb{R}[G]$, which can be found by a simple counting argument. The two bounds are then combined to derive a lower bound of $s$. 

Let us consider the $k_n$-packing number of $G$ with respect to $\ell_1$-norm. Note that it is lower bounded by the $k_n$-covering number of $G$ with respect to $\ell_1$-norm, and the covering number is further lowered by $|G|$ divided by the cardinality of an $\ell_1$-norm ball with radius $k_n$. Hence, there exists a maximal packing $C_G$ with
\begin{equation}\label{eq:converse_a}
%\lvert G\rvert \le \lvert C_G\rvert\sum_{j=0}^{k_n} \binom{n}{j} 
%\implies
\lvert C_G\rvert \ge \frac{\lvert G\rvert}{\sum_{j=0}^{k_n} \binom{n}{j}}
\ge \frac{\lvert G\rvert}{(n+1)^{k_n}}.
% \underbrace{}_{\text{cardinality of an $\ell_1$-norm ball with radius $k_n$}}
\end{equation}

The choice of the subset $G$ is a second key to the proof. Since we are going to upper bound the $2d_n$-packing number with respect to $\ell_\infty$-norm of the image set $\mb{R}[G]$, we select $G$ so that it is \emph{strictly} contained in an $s$-dimensional cube of appropriate side lengths, say, $l_1,l_2,...,l_s$. Then, as $\bm{R}_i\bm{x}$ are integers for all $\bm{x}$ and $i=1,...,s$, the packing number with respect to $\ell_\infty$-norm is upper bounded by
\begin{equation}\label{eq:converse_b}
\prod_{i=1}^s \frac{l_i}{2d_n} = \frac{\prod_{i=1}^s l_i}{\lp2d_n\rp^s}.
\end{equation}

Combining \eqref{eq:converse_a} and \eqref{eq:converse_b}, it can be seen that with $d_n=n^\delta$ and $k_n = n^\kappa$, 
\begin{align}\label{eq:key_ineq}
\frac{\prod_{i=1}^s l_i}{\lp2d_n\rp^s} \ge \frac{\lvert G\rvert}{(n+1)^{k_n}}
\implies
s\lp\sum_{i=1}^s \frac{\log l_i}{s} - \delta \log n -1\rp \ge \log\lvert G\rvert - n^\kappa\log(n+1). 
\end{align}
Hence, to get the desired lower bound on $s$, $l_i \approx \sqrt{n}$ within a poly-logarithm factor and $\lvert G \rvert \approx 2^n$.

The above discussion motivates us to make the following choice of $G$. 
Let $r_i := \frac{1}{2}\lnorm \bm{R}_i\rnorm_{0}$, the number of $1$'s in the $i$-th counting measurement, $i=1,...,s$.  
To this end, let us define ``atypical'' sets to be excluded as follows: for $i=1,2,...,s$, 
\begin{align}
\notag&\text{If $r_i \ge \sqrt{n\log n}$,} \\
\label{eq:B_i_a}&\quad
B_i := \big\{ \bm{x}\in\{0,1\}^n : \lvert \bm{R}_i\bm{x}-r_{i}/2 \rvert \geq \sqrt{r_{i}\log r_i} \big\}.\\
\notag&\text{If $r_i < \sqrt{n\log n}$,} \\
\label{eq:B_i_b}&\quad 
B_i := \emptyset.
\end{align}
%
%
%\begin{equation}\label{eq:choice_bad_set}
%B_i := \big\{ \bm{x}\in\{0,1\}^n : \lvert \bm{R}_i\bm{x}-r_{i}/2 \rvert \geq l_i \big\},
%\end{equation}
%where $r_1,...,r_s$, the levels of \emph{closeness}, will be chosen such that a sufficient number of elements in $\{0,1\}^n$ are included. 
The ``typical'' set to be considered is hence
\begin{equation}\textstyle \label{eq:choice_good_set}
G := \{0,1\}^{n} \setminus B,\ \text{where $B := \bigcup_{i=1}^s B_{i}$.}
\end{equation}

To control the cardinality of $B_i$, let us employ Hoeffiding's inequality as follows: randomize the data vector so that the $n$ elements $X_1,...,X_n$ are now $n$ i.i.d. $\Ber(1/2)$ random variables. In other words, $\bm{X} = [X_1\ X_2\ ... X_n]^\intercal \sim \Unif(\{0,1\}^n)$. As a result, for \eqref{eq:B_i_a}, $\lvert B_i\rvert = 2^n\Pr\big\{ \lvert\bm{R}_i\bm{X} - r_i/2\rvert \ge l_i/2\big\}$, with $l_i = 2\sqrt{r_{i}\log r_i}$. Note that given a pooling vector $\bm{R}_i$, the outcome of the counting measurement, $\bm{R}_i\bm{X}$, is just the sum of $r_i$ i.i.d. $\Ber(1/2)$ random variables. Hence, by Hoeffiding's inequality, with $l_i = 2\sqrt{r_{i}\log r_i}$, 
\[\textstyle
\Pr\big\{ \lvert\bm{R}_i\bm{X} - r_i/2\rvert \ge l_i/2\big\} 
\le 2\e^{-l_i^2/2r_i} 
= 2r_i^{-2} \le \frac{2}{n \log n}.
\]
Consequently, $\forall\,i=1,...,s$, $\lvert B_i\rvert \le 2^n\frac{2}{n\log n}$, and 
\begin{equation}\label{eq:card_G}
\lvert G \rvert \ge 2^n - \sum_{i=1}^s \lvert B_i\rvert 
\ge 2^n\Big( 1- \frac{2s}{n \log n} \Big)
\ge 2^n\Big( 1- \frac{2}{\log n} \Big),
%\lvert B\rvert \le \sum_{i=1}^s \lvert B_i\rvert \le 2^n\sum_{i=1}^s \exp\lp -2r_il_i^2\rp,
\end{equation}
where in the last inequality, we make use of an implicit assumption that $n \ge s$ when $n$ is sufficiently large, due to the achievability part (Lemma~\ref{non_constructive_proof}). 

Let us now turn back to inequality \eqref{eq:key_ineq}. 
%, to upper bound the $d_n$-packing number with respect to $\ell_\infty$-norm of the image set $\mb{R}[G] := \{\mb{R}\bm{x}\,\vert\,\bm{x}\in G\}$. 
The choice of $G$ in \eqref{eq:B_i_a} -- \eqref{eq:choice_good_set} together with the fact that $0\le \bm{R}_i\bm{x}\le r_i$ (since it is the outcome of a counting measurement with $r_i$ items in the pool) ensures that the image set $\mb{R}[G]$ is strictly contained in an $s$-dimensional cube with side lengths not greater than $2\sqrt{n\log n}$. Hence, \eqref{eq:key_ineq} and \eqref{eq:card_G} imply
\begin{align*}
s\Big( \log(2\sqrt{n\log n}) - \delta \log n -1\Big)
\ge \log\lp 2^n\Big( 1- \frac{2}{\log n} \Big)\rp - n^\kappa\log(n+1).
\end{align*}
As $n$ tends to infinity, we conclude that 
\[
\liminf_{n\ra\infty} \frac{s}{n/\log n}\ge \frac{1}{1-2\delta}\ .
\]

The proof for the sparse case (Lemma~\ref{sparse_converse}) largely follows that of the non-sparse case, with slight modification of the definition of the ``atypical'' sets in \eqref{eq:B_i_a} and \eqref{eq:B_i_b}: for $i=1,2,...,s$, the definition of $B_i$ in \eqref{eq:B_i_a} is changed to 
\[\textstyle
%B_i := 
\lbp \bm{x}\in\{0,1\}^n : \lVert\bm{x}\rVert_0 \leq n^{\lambda}, \lvert \bm{P}_i\bm{x}-\frac{p_{i}n^{\lambda}}{n} \rvert \geq \sqrt{6\lambda n^{\lambda}\log n} \rbp.
\]
Accordingly, the ``typical'' set $G$ becomes
\[
\textstyle
G := \{x\in \{0,1\}^{n},\lVert\bm{x}\rVert_0 \leq n^{\lambda}  \} \setminus B,\ \text{where $B := \bigcup_{i=1}^s B_{i}$.}
\]

Chernoff bound is then employed to control the cardinality of $B_i$. Note that the new definition of $B_i$ has an additional \emph{sparsity} constraint $\lVert\bm{x}\rVert_0 \leq n^{\lambda}$. Removing the sparsity constraint, we have a set $\tilde{B}_i$ with cardinality not smaller than that of $B_i$. Now, randomize the data vector so that $X_i\diid\Ber(n^{\lambda-1})$, $i=1,...,n$. 
%As a result, for \eqref{eq:B_i_a}, $\lvert B_i\rvert = 2^n\Pr\big\{ \lvert\bm{Q}_i\bm{X} - r_i/2\rvert \ge l_i/2\big\}$, with $l_i = 2\sqrt{p_{i}\log r_i}$. Note that given a pooling vector $\bm{Q}_i$, the outcome of the counting measurement, $\bm{Q}_i\bm{X}$, is just the sum of $r_i$ i.i.d. $\Ber(1/2)$ random variables. Hence, by Hoeffiding's inequality, with $l_i = 2\sqrt{p_{i}\log r_i}$, 
%Setting $l_i = \sqrt{\frac{p_{i}n^{\lambda}\log n}{n}}$, 
We first calculate 
$\Pr\big\{ \lvert\bm{P}_i\bm{X} - p_{i}n^{\lambda}/n\rvert \ge \sqrt{6\lambda n^{\lambda}\log n} \big\}$, and then relate this quantity to $\lvert \tilde{B}_i\rvert$.

\begin{align*}
    &\Pr\big\{ \lvert\bm{P}_i\bm{X} - \frac{p_{i}n^{\lambda}}{n}\rvert \ge \sqrt{6\lambda n^{\lambda}\log n}\big\}\\ 
 &\le^{(a)} 2\frac{\lp n^{\lambda -1}e^{t}+1-n^{\lambda -1}\rp^{r_i}}{e^{t\lp p_{i}n^{\lambda -1}+\sqrt{6\lambda n^{\lambda}\log n}\rp}} = 2\frac{\lp 1+n^{\lambda -1}\lp e^{t}-1\rp\rp^{p_i}}{e^{t\lp p_{i}n^{\lambda -1}+\sqrt{6\lambda n^{\lambda}\log n}\rp}}\\
 &\leq 2\frac{e^{n^{\lambda -1}\lp e^{t}-1\rp p_i}}{e^{t\lp p_{i}n^{\lambda -1}+\sqrt{6\lambda n^{\lambda}\log n}\rp}}\leq^{(b)} \frac{2}{\lp n^{\lambda}\log n\rp^2}.
\end{align*}

($a$) follows from Chernoff bound. In order to get ($b$), it suffics to choose $t$ such that  
\begin{align}\label{satis_1}
    &t\lp p_{i}n^{\lambda -1}+\sqrt{6\lambda n^{\lambda}\log n}\rp - n^{\lambda -1}\lp e^{t}-1\rp p_i\\
    &\geq 2\lambda \log(n) + 2\log\lp\log(n)\rp
    \label{satis_2}
\end{align}

Then we have

\begin{align*}
    &\eqref{satis_1}\geq t\lp p_{i}n^{\lambda -1}+\sqrt{6\lambda n^{\lambda}\log n}\rp - n^{\lambda -1}\lp e^{t}-1\rp p_i\\
    &=^{(a)} \ln\lp 1+\frac{\sqrt{6\lambda n^{\lambda}\log n}}{n^{\lambda -1}p_{i}}\rp \lp p_{i}n^{\lambda -1}+\sqrt{6\lambda n^{\lambda}\log n}\rp - \sqrt{6\lambda n^{\lambda}\log n}:=f(p_{i})\\ 
    &\geq^{(b)} \ln\lp 1+\frac{\sqrt{6\lambda n^{\lambda}\log n}}{n^{\lambda}}\rp \lp n^{\lambda}+\sqrt{6\lambda n^{\lambda}\log n}\rp- \sqrt{6\lambda n^{\lambda}\log n}\\
    &\geq^{(c)} \lp \frac{\sqrt{6\lambda n^{\lambda}\log n}}{n^{\lambda}} - \frac{6\lambda n^{\lambda}\log n}{2n^{2\lambda}}\rp\lp n^{\lambda}+\sqrt{6\lambda n^{\lambda}\log n}\rp - \sqrt{6\lambda n^{\lambda}\log n}\\
    &= \frac{1}{2}\frac{6\lambda n^{\lambda}\log n}{n^{\lambda}}\lp 1-\frac{\sqrt{6\lambda n^{\lambda}\log n}}{n^{\lambda}}\rp
    \geq 2\lambda \log n + 2\log\lp\log n\rp
\end{align*}

($a$) follows from choose $t = \ln\lp 1+\frac{\sqrt{6\lambda n^{\lambda}\log n}}{n^{\lambda}}\rp$.($b$) follows from the the fact that $f(p_{i})$ is a decreasing function of $p_{i}$ and $1\leq p_{i}\leq n$. ($c$) follows from $\ln(1+x)\geq x-\frac{x^2}{2}$.

Consequently, $\forall\,i=1,...,s$, $\lvert B_i\rvert \le^{(a)} {n\choose n^{\lambda}}n^{\lambda}\frac{2}{n^{2\lambda} (\log n)^{2}}$, and 
\begin{align}\label{sparse_set_size}
\lvert G \rvert &\ge {n\choose n^{\lambda}} - \sum_{i=1}^s \lvert B_i\rvert 
\ge {n\choose n^{\lambda}}\Big( 1- \frac{2s}{n^{\lambda} (\log n)^{2}} \Big)\\
&\ge^{(b)} n^{(1-\lambda)n^{\lambda}}\Big( 1- \Theta(\frac{1}{\log n}) \Big),
%\lvert B\rvert \le \sum_{i=1}^s \lvert B_i\rvert \le 2^n\sum_{i=1}^s \exp\lp -2p_il_i^2\rp,
\end{align}

($a$) follows from the fact that for all $x\in \{0,1\}^{n},\lVert\bm{x}\rVert_0 \leq n^{\lambda}$, those $\lVert\bm{x}\rVert_0 = n^{\lambda}$ has the smallest probability $\lp\frac{n^\lambda}{n}\rp^{n^\lambda}\lp 1-\frac{n^\lambda}{n}\rp^{n-n^{\lambda}}$, and ${n\choose n^{\lambda}}\lp\frac{n^\lambda}{n}\rp^{n^\lambda}\lp 1-\frac{n^\lambda}{n}\rp^{n-n^{\lambda}}\geq \frac{1}{n^\lambda}$.
where in ($b$), we make use of an implicit assumption that $s = O(n^{\lambda}\log n)$ when $n$ is sufficiently large, due to the achievability part (Lemma~\ref{sparse_achievability}), and the fact that ${n\choose n^{\lambda}}\geq n^{(1-\lambda)n^{\lambda}}$.

Finally, combine \eqref{eq:key_ineq},\eqref{sparse_set_size}, we get

\begin{align*}
s\Big( \log(2\sqrt{6\lambda n^{\lambda}\log n}) - \delta \log n -1\Big)
\ge \log\lp n^{(1-\lambda)n^{\lambda}}\Big( 1- \Theta(\frac{1}{\log n}) \Big)\rp - n^\kappa\log(n+1).
\end{align*}
As $n$ tends to infinity, we conclude that 
\[
s \ge
\begin{cases}
                \frac{2(1-\lambda)}{\lambda-2\delta}n^\lambda,  &2\delta<\kappa   \\[1ex]
                \frac{2(1-\lambda)}{\lambda-2\delta}n^{\lambda}\log n,  &2\delta=\kappa
\end{cases}
\]

%\[\textstyle
%\forall\,\bm{x} \in G,\ \bm{Q}_i\bm{x} \in \big( \frac{p_i}{2}-l_i, \frac{p_i}{2}+l_i \big).
%\]
%As $\bm{Q}_i\bm{x}$ are integers for all $\bm{x}$ and $i=1,...,s$, the packing number with respect to $\ell_\infty$-norm is upper bounded by
%\begin{equation}
%\prod_{i=1}^s \frac{2l_i}{d_n} = \frac{2^s\prod_{i=1}^s l_i}{d_n^s}.
%\end{equation}

%\begin{equation}
%\begin{cases}
%\text{If $p_i \ge 2\sqrt{n\log n}$,} \\
%\quad
%B_i := \big\{ \bm{x}\in\{0,1\}^n : \lvert \bm{P}_i\bm{x}-p_{i}/2 \rvert \geq 2\sqrt{p_{i}\log p_i} \big\}.\\
%\text{If $p_i < 2\sqrt{n\log n}$,} \\
%\quad 
%B_i := \emptyset.
%\end{cases}
%\end{equation}
%\begin{equation}
%B_i :=
%\begin{cases}
%\big\{ \bm{x}\in\{0,1\}^n : \lvert \bm{P}_i\bm{x}-\frac{p_{i}}{2} \rvert \geq 2\sqrt{p_{i}\log p_i} \big\}, &p_i \ge 2\sqrt{n\log n}\\
%\emptyset, &\text{otherwise}.
%\end{cases}
%\end{equation}
%Roughly speaking, $B_i$ denotes the set of data vectors $\bm{x}$ that do not concentrate for specific measurement $p_i$.
%The ``good'' set to be considered is hence
% \begin{equation}\textstyle
%G := \{0,1\} \setminus B,\ \text{where $B := \bigcup_{i=1}^s B_{i}$.}
%\end{equation}
%The whole proof is base on counting, the key ingredient is that given any measurement $\bm{\mb{M}}$, for most of the $\bm{x}\in \{0,1\}^n$, $y = \bm{Mx}$ will concentrate in $[\frac{m_{i}}{2}-\sqrt{m_{i}\log(m_i)}, \frac{m_{i}}{2}+\sqrt{m_{i}\log(m_i)}]$, where $m_{i} = \lba M\rba _{0}$.

\section{Discussion, Conclusion, and Future work}\label{sec:conclusion}
\subsection{Discussion}
{In section~\ref{adaptive_embedding}, we focus on applying our embedding method to the specific noisy SCQGT problem. However, its high-level idea has potential in constructing deterministic and non-adaptive measurement algorithms for explicit construction in other problems with a sparse structure. We describe the embedding principle as an extension of our embedding method in the construction of the noisy SCQGT problem.

There are two steps in our embedding principle, which can be applied to construct deterministic and non-adaptive pooling algorithms for general problems with a sparse structure. First, we consider the same problem but without a sparse structure. Typically, the difficulty would be dramatically reduced. For example, a noisy, compressed sensing problem without sparse constraints reduce to a linear regression problem. Therefore we can get deterministic and non-adaptive constructions $\mb{M}$ more easily. Second, we tried to embed many small-versioned $\mb{M}$ into the final pooling matrix $\mb{M}_{s}$ for the original problem with sparse constraints so that whatever the support set of the hidden data is, the corresponding column-submatrix of $\mb{M}_{s}$ always contain a submatrix similar to $\mb{M}$. Consequently, $\mb{M}_{s}$ is a good pooling matrix for the original problem with sparse constraints.
}
\subsection{Conclusion}
{
In this work, we investigate combinatorial quantitative group testing (CQGT) problems, without or with a sparsity constraint, under adversarial but bounded noise. In particular, we assume that the noise vector is bounded by $n^\delta$ in $l_{\infty}$ norm and aim to reconstruct $n^\kappa$ out of $n$ bits. It falls into a broader class of problems where the goal is to recover a signal from group-wise information.

For CQGT, first, we characterize the fundamental limit, which turns out to be $\Theta\lp \frac{1}{1-2\delta}\frac{n}{\log\lp n\rp} \rp$. Then, we construct an optimal deterministic pooling matrix with an efficient $o\lp n^{1+\epsilon}\rp, \forall\epsilon >0$ decoding algorithm. Although we are focused on the binary-valued signals in this work, with a slight adjustment, our results can be extended to general integer-valued signals.

For SCQGT with sparsity level $n^\lambda$, we again characterize the fundamental limit on the pooling complexity, which turns out to be $\Theta\lp\frac{1-\lambda}{\lambda-2\delta} n^{\lambda}\rp$. With two kinds of matrices-the Hadamard matrix and the adjacency matrix of bipartite expanders-as building blocks, we construct a near-optimal pooling matrix with pooling complexity $n^{\lambda} \lp\frac{\lambda\log(n)^{2}}{\epsilon}\rp^{3}4^{\sqrt{3\lambda \log\lp n\rp\log \lp\frac{\lambda\log(n)^{2}}{\epsilon}\rp}}$ and an efficient $O\lp n^{1+\epsilon}\rp, \forall \epsilon >0$ decoding algorithm. It is worth noting that this SCQGT sensing matrix can also distinguish sparse real-valued signals and hence can be applied to robust compressed sensing. The underlying embedding principle of constructed SCQGT algorithm has potential for other problems with sparse structures.

Finally, we emphasize that all pooling algorithms proposed in this work are combinatorial and non-adaptive, and hence they are ready for practical uses. Since the setting is combinatorial, the constructions can provide a deterministic guarantee and reduce the decoding complexity of time and space by leveraging its specific structure. Moreover, since they are non-adaptive, all pools can be conducted simultaneously, and further speed up is possible if parallelization is done properly.

\subsection{Future works}
There are several interesting directions.
\begin{enumerate}
    \item Extending our embedding idea to other problems with sparse structure such as robust compressed sensing.
    \item Time-varying data: In this work, the data we aim to recover is fixed and does not change between pools. Meanwhile, in many real applications such as group testing which aims to screen for the presence of virus of Covid-19, healthy individuals might be infected between pools, and thus the data we aim to recover change over time. Hence, it is an interesting direction for future works.   
    \item Correlated data: In this work, we focus on the combinatorial guarantee and hence can be applied to any data distribution. In practice, however, we usually deal with more constrained data distribution, and thus smaller pooling complexity is possible. For example, in the application of screening for the presence of virus of Covid-19, the data we aim to recover is highly correlated since there exist underlying social structures.
    \item Optimal non-adaptive measurement plan for SCQGT: The fundamental limit of the pooling complexity for SCQGT, by the previous discussion, is $\Theta\lp n^{\lambda}\rp$. To date, all existing explicit non-adaptive constructions are sub optimal even for noiseless SCQGT. They only leverage sparse structures. But one must consider its arithmetic progression structures in order to reach its fundamental limit. 
\end{enumerate}

}

\bibliographystyle{IEEEtran}
\bibliography{Ref.bib}

%\newpage
%\onecolumn

%\appendix
%\section{Appendix}\label{sec:proofs}
%\input{\srcpath sec_proofs.tex}

%\section*{Acknowledgment}

%%%%%%
%% To balance the columns at the last page of the paper use this
%% command:
%%
%\enlargethispage{-1.2cm} 
%%
%% If the balancing should occur in the middle of the references, use
%% the following trigger:
%%
%\IEEEtriggeratref{3}
%%
%% which triggers a \newpage (i.e., new column) just before the given
%% reference number. Note that you need to adapt this if you modify
%% the paper.  The "triggered" command can be changed if desired:
%%
%\IEEEtriggercmd{\enlargethispage{-20cm}}
%%
%%%%%%

%%%%%%
%% References:
%% We recommend the usage of BibTeX:
%%

\end{document}